\documentclass[aps,reprint,nofootinbib,tightenlines]{revtex4-1}
\usepackage{amsfonts}
\usepackage{amsmath,amssymb,amsthm}
\usepackage[pdftex]{graphicx}
\usepackage[pdftex,colorlinks]{hyperref}
\usepackage{mathrsfs}
\usepackage[normalem]{ulem}

\begin{document}

\title{Nonequilibrium Dynamics of Charged Particles in an Electromagnetic Field:\\
Causal and Stable Dynamics from $1/c$ Expansion of QED}
\author{C. H. Fleming}
\email{hfleming@physics.umd.edu}
\affiliation{Joint Quantum Institute and Department of Physics, University of Maryland,
College Park, Maryland 20742}
\author{P. R. Johnson}
\email{pjohnson@american.edu}
\affiliation{Department of Physics, American University, Washington D.C. 20016}
\author{B. L. Hu}
\email{blhu@physics.umd.edu}
\affiliation{Joint Quantum Institute and Maryland Center for Fundamental Physics, University of Maryland,
College Park, Maryland 20742}

\begin{abstract}
We derive from a microscopic Hamiltonian a set of stochastic equations of
motion for a system of spinless charged particles in an electromagnetic (EM)
field based on a consistent application of a dimensionful $1/c$ expansion of
quantum electrodynamics (QED). All relativistic corrections up to order $%
1/c^3$ are captured by the dynamics, which includes electrostatic
interactions (Coulomb), magnetostatic backreaction (Biot-Savart),
dissipative backreaction (Abraham-Lorentz) and quantum field fluctuations at
zero and finite temperatures. With self-consistent backreaction of the EM
field included we show that this approach yields causal and runaway-free
equations of motion, provides new insights into charged particle
backreaction, and naturally leads to equations consistent with the
(classical) Darwin Hamiltonian and has quantum operator ordering consistent
with the Breit Hamiltonian. To order $1/c^3$ the approach leads to a
nonstandard mass renormalization which is associated with magnetostatic
self-interactions, and no cutoff is required to prevent runaways. Our new
results also show that the pathologies of the standard Abraham-Lorentz
equations can be seen as a consequence of applying an inconsistent (i.e.
incomplete, mixed-order) expansion in $1/c$, if, from the start, the
analysis is viewed as generating a low-energy effective theory rather than
an exact solution. Finally, we show that the $1/c$ expansion within a
Hamiltonian framework yields well-behaved noise and dissipation, in addition
to the multiple-particle interactions.
\end{abstract}

\date{\today}
\maketitle


\section{Introduction}

The backreaction of a charged particle interacting with the electromagnetic
field involves a number of famous problems including acausality, in the form
of pre-acceleration, and runaway solutions to the Abraham-Lorentz equation 
\cite{Jackson98}. A number of approaches to resolving these problems have
been developed, including replacing point particles with extended objects 
\cite{Yaghjian92,Rohrlich97,Medina06}, treating the electromagnetic field
interaction perturbatively and truncating at a specific order \cite%
{BaroneCaldeira91}, or replacing time-local differential equations of motion
with nonlocal integro-differential equations of motion \cite%
{FordOconnell88,Ford91}. Yet many problems related to the backreaction
remain poorly understood or unresolved. 
The interaction of a local (particle) degree of freedom with a nonlocal
environment (a field) can also result in surprising dynamics, for example
the vanishing of the radiation-reaction force on a uniformly accelerating
charge.

To find the backreaction on a single charged particle within an open systems
framework, one integrates out the field degrees of freedom following
well-known procedures \cite{FordOconnell88}. Starting from a nonrelativistic
Hamiltonian of a charged particle coupled to an electromagnetic field, the
procedure gives the Abraham-Lorentz Langevin equation for a \emph{%
structureless} point charge coupled to the electromagnetic field \cite%
{Dalibard82,FordOconnell88,BaroneCaldeira91}. Including driving by external
forces, the stochastic equations of motion are 
\begin{equation}
m\,\ddot{\mathbf{x}}(t)=\mathbf{F}_{\mathrm{ext}}(t)+\underbrace{%
2\,e^{2}\,\gamma _{0}\,\dddot{\mathbf{x}}(t)}_{\mathrm{backreaction}}-%
\underbrace{e\,\dot{\boldsymbol{\xi }}(t)}_{\mathrm{noise}}\,,
\label{eq:ALL1}
\end{equation}%
where $\gamma _{0}=1/12\pi \varepsilon _{0}c^{3}$ and $\,\dot{\boldsymbol{%
\xi }}(t)$ is quantum-field induced noise. These equations of motion can
describe either classical trajectories or Heisenberg-picture operators. In
the latter case, the induced noise is operator valued and has both a
complex-valued noise correlation and a state-independent commutator. As
noted above, Eq. (\ref{eq:ALL1}) is not manifestly causal and exhibits
runaway solutions. The result in Eq. (\ref{eq:ALL1}) is essentially
equivalent to backreaction with a supra-Ohmic bath in quantum Brownian
motion (QBM) \cite{BaroneCaldeira91,HuMatacz94}. However, relativistic QED\
has nonlinear coupling of system and environment, whereas QBM has bilinear
coupling. Consequently, the physics of QED is far more subtle and complex.

In this paper, we construct an alternative derivation of the stochastic
equations of motion for a system of spinless charged particles based upon a
consistent application of a dimensionful $1/c$ expansion. We show that this
approach yields causal (i.e., no-preacceleration) and stable (i.e., runaway
free) backreaction, provides new insights into previous results, and
naturally leads to equations with a form consistent with the (classical)
Darwin Hamiltonian and operator ordering consistent with the Breit
Hamiltonian. The role of magnetostatic interactions in both mass
renormalization and as a form of dissipationless backreaction is also
revealed.\footnote{%
By electrostatic and magnetostatic we refer to the lowest-order (in $1/c$)
electric and magnetic fields. These fields are instantaneous and accompany
even static charge and current distributions, though our system is not
static. The higher-order relativistic contributions contain retarded and
radiative effects.} For a single particle, our damping admits correspondence
with the casual and stable equations of motion obtained previously by Ford
and O'Connell \cite{Ford91}.

To order $1/c^{3}$ the approach leads to a nonstandard mass renormalization
which is associated with magnetostatic self-interactions, and (to this
order) no cutoff is required to prevent runaways. An important question is
whether the $1/c$ expansion automatically generates consistent low-energy
behavior that is always insensitive to particle structure. Although it will
be complicated, our method provides a systematic method to consistently
extend the analysis to higher orders for a closer investigation of this
question. This is important because the standard analysis giving the
Abraham-Lorentz equation, viewed perturbatively in $1/c,$ is of mixed order.
Our results show that inclusion of incomplete higher-order information (that
is, some but not all terms beyond $1/c^{3}$) can be identified as the cause
of the Abraham-Lorentz equation's perturbative instability, from an
effective theory perspective. Equivalently, our results indicate that in
terms of a $1/c$ expansion the dipole approximation is inconsistently
applied in the standard derivations, and that if all terms present in the
standard calculation are to be included, then additional multipole field
corrections will be also required to the same order. 

Our analysis of the single particle dynamics also highlights a subtle but
important connection between dissipation and noise. There are two standard
calculations for radiation reaction which arise from different choices of
gauge: Coulomb or electric dipole. We show that the former (see, e.g., \cite%
{Dalibard82} for an example) has a problematic description of noise: it is
not manifestly thermal in the sense that the stochastic process present in
the Langevin equation is not sampled from an independent thermal
distribution constructed from the field's Hamiltonian acting as a reservoir.
The latter (see, e.g., \cite{BaroneCaldeira91} for an example) has ordinary
noise but involves integration kernels which are approximately given by the
second derivatives of a delta function, making them more pathological than
the those in the Coulomb gauge. In other words, either the noise or
dissipation has undesirable features in the usual derivations. We show that
the $1/c$ expansion in the Coulomb gauge, with a correct delineation of
system and environment, naturally avoids these problems, yielding
well-behaved noise and dissipation in addition to the multiparticle
interactions within a unified framework.

In Sec.~\ref{sec:QED}, we first analyze nonrelativistic quantum particles in
the electromagnetic field starting from the usual Hamiltonian but instead of
taking the dipole approximation we expand in orders of powers in $1/c$.
Continuing with the usual calculation we reproduce the standard results,
but eventually see that they are \textquotedblleft mixed order\textquotedblright\ in
powers of $1/c$. In Sec.~\ref{sec:EQED}, we work from the Coulomb gauge and
throughout the calculation we consistently remove all terms in the
open-system dynamics beyond $\mathcal{O}\!\left( 1/c^{3}\right) $. The usual
formulations can in principle be applied but it is not as transparent and is
more complicated than necessary, as unneeded higher order terms are first
inconsistently kept and only later removed. Section~\ref{sec:EQED} also
shows that the order $1/c^{3}$ equations of motion yield stable backreaction
and self-consistent noise. Section \ref{sec:Discussion} discusses the regime
of validity of our analysis and discusses future directions. In App.~\ref%
{sec:QBM}, we review the derivation for backreaction in QBM which parallels
our analysis in many respects. Other characteristics of the QBM model which
are used here, such as the renormalization and integration kernels, are more
thoroughly discussed in \cite{QBM}. In Appendix \ref{sec:Explicit} we solve
the field equations of motion sourced by the particles, without invoking the
dipole approximation.


\section{Standard ``Nonrelativistic'' Radiation Reaction}

\label{sec:QED}

We first review the derivation of the standard Hamiltonian for the motion of
a system of N particles, while also defining the notation we will use in the
rest of the paper. We start with the Lagrangian $\mathcal{L}_{\mathrm{sys}}$
for the motion of the system defined as the collection of particles $%
j=1,...,N$ with mass $m_{j}$ and charge $e_{j}$ for the $j^{th}$ particle
and the Lagrangian $\mathcal{L}_{\mathrm{env}}$ for an electro ($\mathbf{E}$%
)- magnetic ($\mathbf{B}$) (EM) field acting as its environment: 
\begin{align}
\mathcal{L}_{\mathrm{QED}}& =\mathcal{L}_{\mathrm{sys}}+\mathcal{L}_{\mathrm{%
int}}+\mathcal{L}_{\mathrm{env}}\,, \\
\mathcal{L}_{\mathrm{sys}}& \equiv \sum_{j}\frac{1}{2}m_{j}\dot{\mathbf{x}}%
_{j}^{2}\,, \\
\mathcal{L}_{\mathrm{env}}& =\int \!d^{3}x^{\prime }\frac{\varepsilon _{0}}{2%
}\left\{ \mathbf{E}(\mathbf{x}^{\prime })^{2}-c^{2}\,\mathbf{B}(\mathbf{x}%
^{\prime })^{2}\right\} \,.
\end{align}%
The interaction between the system of charged particles and the EM field is
given by 
\begin{align}
\mathcal{L}_{\mathrm{int}}& =\int \!d^{3}x^{\prime }\left\{ \mathbf{J}(%
\mathbf{x},\mathbf{x}^{\prime })\cdot \mathbf{A}(\mathbf{x}^{\prime })-\rho (%
\mathbf{x},\mathbf{x}^{\prime })\,\phi (\mathbf{x}^{\prime })\right\} \,, \\
& =\sum_{j}e_{j}\left\{ \dot{\mathbf{x}}_{j}\cdot \mathbf{A}(\mathbf{x}%
_{j})-\phi (\mathbf{x}_{j})\right\} \,,
\end{align}%
where $\rho $ and $\mathbf{J}$ are the charge and current density coupled to
the scalar and vector potentials $\phi ,\mathbf{A}$ respectively which are
related to the electromagnetic fields by 
\begin{align}
\mathbf{E}(\mathbf{x})& =-\boldsymbol{\nabla }\phi (\mathbf{x})-\frac{%
\partial }{\partial t}\mathbf{A}(\mathbf{x})\,, \\
\mathbf{B}(\mathbf{x})& =\boldsymbol{\nabla }\!\times \!\mathbf{A}(\mathbf{x}%
)\,.
\end{align}%
Expressing the vector potential in terms of the spatial Fourier modes with
wavevector $\mathbf{k}$ and polarization $\boldsymbol{\epsilon }_{k}$ gives 
\begin{align}
\mathbf{A}(\mathbf{x})& =\frac{1}{(2\pi )^{3/2}}\int \!d^{3}k\sum_{%
\boldsymbol{\epsilon }_{k}}\mathbf{A}_{\mathbf{k},\boldsymbol{\epsilon }%
_{k}}(\mathbf{x})\,,  \label{eq:Ak} \\
\mathbf{A}_{\mathbf{k},\boldsymbol{\epsilon }_{k}}(\mathbf{x})& =\frac{%
\boldsymbol{\epsilon }_{k}}{\sqrt{2\varepsilon _{0}\omega _{k}}}\left\{
e^{+\imath \mathbf{k}\cdot \mathbf{x}}\,\mathbf{a}_{\mathbf{k},\boldsymbol{%
\epsilon }_{k}}+e^{-\imath \mathbf{k}\cdot \mathbf{x}}\,\mathbf{a}_{\mathbf{k%
},\boldsymbol{\epsilon }_{k}}^{\dagger }\right\} ,  \label{eq:Ak2}
\end{align}%
where $\omega _{k}=c\,k$. To satisfy the commutation relations, the
conjugate momentum of the field is then given by 
\begin{align}
\boldsymbol{\mbox{\Large$\pi$}}(\mathbf{x})& =\frac{1}{(2\pi )^{3/2}}\int
\!d^{3}k\sum_{\boldsymbol{\epsilon }_{k}}\boldsymbol{\mbox{\Large$\pi$}}_{%
\mathbf{k},\boldsymbol{\epsilon }_{k}}(\mathbf{x})\,,  \label{eq:Pik} \\
\boldsymbol{\mbox{\Large$\pi$}}_{\mathbf{k},\boldsymbol{\epsilon }_{k}}(%
\mathbf{x})& =\frac{-\imath \,\boldsymbol{\epsilon }_{k}}{\sqrt{%
2/\varepsilon _{0}\omega _{k}}}\left\{ e^{+\imath \mathbf{k}\cdot \mathbf{x}%
}\,\mathbf{a}_{\mathbf{k},\boldsymbol{\epsilon }_{k}}-e^{-\imath \mathbf{k}%
\cdot \mathbf{x}}\,\mathbf{a}_{\mathbf{k},\boldsymbol{\epsilon }%
_{k}}^{\dagger }\right\} \,.
\end{align}%
To systematically treat the position dependence of the coupling, we write%
\begin{align}
\mathbf{A}_{\mathbf{k},\boldsymbol{\epsilon }_{k}}(\mathbf{x})& =\cos (%
\mathbf{k}\cdot \mathbf{x})\,\mathbf{A}_{\mathbf{k},\boldsymbol{\epsilon }%
_{k}}-\sin (\mathbf{k}\cdot \mathbf{x})\frac{\boldsymbol{\mbox{\Large$\pi$}}%
_{\mathbf{k},\boldsymbol{\epsilon }_{k}}}{\varepsilon _{0}\omega _{k}}\,, \\
\boldsymbol{\mbox{\Large$\pi$}}_{\mathbf{k},\boldsymbol{\epsilon }_{k}}(%
\mathbf{x})& =\cos (\mathbf{k}\cdot \mathbf{x})\,\boldsymbol{%
\mbox{\Large$\pi$}}_{\mathbf{k},\boldsymbol{\epsilon }_{k}}+\sin (\mathbf{k}%
\cdot \mathbf{x})\,\varepsilon _{0}\omega _{k}\,\mathbf{A}_{\mathbf{k},%
\boldsymbol{\epsilon }_{k}}\,,
\end{align}%
where%
\begin{align}
\mathbf{A}_{\mathbf{k},\boldsymbol{\epsilon }_{k}}& \equiv \mathbf{A}_{%
\mathbf{k},\boldsymbol{\epsilon }_{k}}(\mathbf{0})\,, \\
\boldsymbol{\mbox{\Large$\pi$}}_{\mathbf{k},\boldsymbol{\epsilon }_{k}}&
\equiv \boldsymbol{\mbox{\Large$\pi$}}_{\mathbf{k},\boldsymbol{\epsilon }%
_{k}}(\mathbf{0})\,,
\end{align}%
are $\mathbf{x}$-independent field operators that more closely correspond to
the \textquotedblleft positions\textquotedblright\ and \textquotedblleft
momenta\textquotedblright , $q_{\mathbf{k}}$ and $\pi _{\mathbf{k}},$ of the
reservoir oscillators for QBM, defined in App.~\ref{sec:QBM}.

\subsection{Coulomb-Gauge Hamiltonian}

\label{sec:QEDv} Choosing the Coulomb gauge $\boldsymbol{\nabla }\cdot 
\mathbf{A}=0$, the scalar potential $\phi $ becomes the nondynamical Coulomb
potential and the vector potential $\mathbf{A}$ is purely transverse, i.e. $%
\mathbf{k}\cdot \boldsymbol{\epsilon }_{k}=0$. The system plus environment
Hamiltonian may then be expressed as 
\begin{align}
\mathbf{H}_{\mathrm{QED}}^{A}& =\sum_{j}\left\{ \frac{\left[ \mathbf{p}%
_{j}-e_{j}\,\mathbf{A}(\mathbf{x}_{j})\right] ^{2}\!\!}{2m_{j}}+e_{j}\,\phi (%
\mathbf{x}_{j})\right\} +\mathbf{H}_{\mathrm{field}}\,,  \label{eq:HQED1} \\
\mathbf{H}_{\mathrm{field}}& \equiv \int \! d^{3} x^{\prime }\frac{1}{2}%
\left\{ \varepsilon _{0}^{-1}\boldsymbol{\mbox{\Large$\pi$}}(\mathbf{x}%
^{\prime})^2+\varepsilon _{0}c^{2}\left[ \boldsymbol{\nabla }\!\times \!%
\mathbf{A}(\mathbf{x}^{\prime })\right] ^{2}\right\} \,.
\end{align}%
%
%
%
%
%
Expanding in field modes, the Hamiltonian is 
\begin{equation}
\mathbf{H}_{\mathrm{field}}\equiv \int \! d^{3}k\sum_{\boldsymbol{\epsilon }%
_{k}}\frac{1}{2}\left\{ \varepsilon _{0}^{-1}\boldsymbol{\mbox{\Large$\pi$}}%
_{\mathbf{k},\boldsymbol{\epsilon }_{k}}^{2}\!\!+\varepsilon _{0}\omega
_{k}^{2}\,\mathbf{A}_{\mathbf{k},\boldsymbol{\epsilon }_{k}}^{2}\right\} .
\label{eq:Hfield1}
\end{equation}

The Heisenberg equations of motion for the system as driven by the field are
then 
\begin{align}
\dot{\mathbf{x}}_{j}& =\frac{\mathbf{p}_{j}}{m_{j}}-\frac{e_{j}}{m_{j}}%
\mathbf{A}(\mathbf{x}_{j})\,,  \label{eq:Lv1} \\
\dot{\mathbf{p}}_{j}& =-e_{j}\boldsymbol{\nabla }_{\!\!j}\phi (\mathbf{x}%
_{j})+\frac{1}{2m_{j}}\boldsymbol{\nabla }_{\!\!j}\left[ \mathbf{p}%
_{j}-e_{j}\,\mathbf{A}(\mathbf{x}_{j})\right] ^{2}\,,  \label{eq:Lv2}
\end{align}%
and the Heisenberg equations of motion for the environment as driven by the
system are given by 
\begin{align}
\dot{\mathbf{A}}_{\mathbf{k},\boldsymbol{\epsilon }_{k}}& =\frac{1}{%
\varepsilon _{0}}\boldsymbol{\mbox{\Large$\pi$}}_{\mathbf{k},\boldsymbol{%
\epsilon }_{k}}+\frac{1}{\varepsilon _{0}\omega _{k}}\sum_{j}\frac{e_{j}}{2}%
\left\{ \sin (\mathbf{k}\cdot \mathbf{x}_{j}),\dot{\mathbf{x}}_{j}\right\}
\,, \\
\dot{\boldsymbol{\mbox{\Large$\pi$}}}_{\mathbf{k},\boldsymbol{\epsilon }%
_{k}}& =-\varepsilon _{0}\omega _{k}^{2}\,\mathbf{A}_{\mathbf{k},\boldsymbol{%
\epsilon }_{k}}+\sum_{j}\frac{e_{j}}{2}\left\{ \cos (\mathbf{k}\cdot \mathbf{%
x}_{j}),\dot{\mathbf{x}}_{j}\right\} \,,
\end{align}
where $\left\{ A,B\right\} =A\,B+B\,A$ is the anticommutator. From App.~\ref%
{sec:Explicit}, the driven solution can be expressed in the
manifestly-Hermitian form 
\begin{equation}
\mathbf{A}(\mathbf{x}_{i},t)=\boldsymbol{\xi }_{i}^{A}(t)-\sum_{j}e_{j}\left%
\{ (\boldsymbol{\mu }_{ij}^{A}\ast \dot{\mathbf{x}}_{j})(t)+(\boldsymbol{\mu 
}_{ij}^{A}\ast \dot{\mathbf{x}}_{j})^{\dagger }(t)\right\} \,,
\label{eq:A(v)}
\end{equation}
with the convolutions defined as 
\begin{equation*}
(\boldsymbol{\mu }_{ij}^{A}\ast \dot{\mathbf{x}}_{j})(t)
=\int_{0}^{t}\!\!dt^{\prime }\boldsymbol{\mu }_{ij}^{A}[\mathbf{x}_{i}(t),%
\mathbf{x}_{j}(t^{\prime });t\!-\!t^{\prime }]\,\dot{\mathbf{x}}%
_{j}(t^{\prime })\,,
\end{equation*}%
where the dissipation kernel $\boldsymbol{\mu }^{A}$ is defined exactly in
App.~\ref{sec:Explicit} and resolved (to lowest orders in $1/c$) in Sec.~\ref%
{sec:gamma}. The dissipation-kernel integral arises from the system driving
the field, whereas the operator $\boldsymbol{\xi }_{i}^{A}(t)$ corresponds
to the homogeneous evolution of the field.

Multiparticle Langevin equations of motion can then be obtained by
substituting Eq.~\eqref{eq:A(v)} for $\mathbf{A}$ back into the system
equations of motion \eqref{eq:Lv1}-\eqref{eq:Lv2}. We will not write down
the multiparticle equations in this approach, however, since they are more
cumbersome than the form we obtain in Sec.~\ref{sec:EQED}. We analyze the
single-particle case in Sec.~\ref{sec:Standard}. This gives the
Abraham-Lorentz Langevin result in \cite{Dalibard82}.

Although the Langevin equation derived from Eq.~\eqref{eq:A(v)} is valid as
an equation of motion for the Heisenberg operators, proper global boundary
conditions must be applied so that the homogeneous-evolution operator $%
\boldsymbol{\xi }_{i}^{A}(t)$ can be given the interpretation of
independently-sampled thermal noise. (For the simpler case of QBM these
details are described in App.~\ref{sec:factor}-\ref{sec:FDR}.) To obtain
statistically-independent thermal noise, the Langevin equation must be
derived assuming an initially factorized system and environment state $%
\boldsymbol{\rho }_{\mathrm{sys+env}}=\boldsymbol{\rho }_{\mathrm{sys}%
}\otimes \boldsymbol{\rho }_{\mathrm{env}}$ with the environment initially
in its equilibrium state. (Of course $\boldsymbol{\rho }_{\mathrm{sys+env}}$
is not an equilibrium state at the initial time, only $\boldsymbol{\rho }_{%
\mathrm{env}}$.) However the field equations in Eq.~\eqref{eq:A(v)} are
expressed in terms of the particle velocity $\dot{\mathbf{x}}_{j}(t)$ rather
than the canonical momentum $\mathbf{p}_{j}(t)$ and this implies that the
initial state of the system plus environment cannot immediately be placed
into the standard form with respect to the Langevin equation.

To see this, note that for a Langevin equation sourced with $\dot{\mathbf{x}}%
_{j}(t)$ we must supply initial data $\mathbf{x}_{j}(0)$, $\dot{\mathbf{x}}%
_{j}(0)$, etc., for the particle that is independent of (uncorrelated with)
the initial state of the environment. Most naively, such an initial state
would have the factorized form%
\begin{equation*}
\boldsymbol{\rho }_{\mathrm{sys+env}}=\boldsymbol{\rho }_{\mathrm{sys}}(%
\mathbf{x},\dot{\mathbf{x}})\otimes \boldsymbol{\rho }_{\mathrm{env}}(%
\mathbf{A},\boldsymbol{\mbox{\Large$\pi$}})\,.
\end{equation*}%
However, because the velocity $\dot{\mathbf{x}}_{j}(t)$ does not commute
with the canonical momentum $\boldsymbol{\mbox{\Large$\pi$},}$ such a state
with $\boldsymbol{\rho }_{\mathrm{env}}(\mathbf{A},\boldsymbol{%
\mbox{\Large$\pi$}})$ cannot represent a (field) equilibrium state with
respect to Hamiltonian \eqref{eq:Hfield1}. It is also likely that such a
\textquotedblleft factorized\textquotedblright\ initial state is not even a
proper density matrix as the product of two noncommuting positive-definite
matrices is not necessarily positive definite. A nontrivial initial state
that did take the above form would necessarily involve, therefore, an
initial \emph{nonequilibrium} state of the environment that implicitly
depends on $\boldsymbol{\rho }_{\mathrm{sys}}(\mathbf{x},\dot{\mathbf{x}}).$
The operator-valued quantum noise $\boldsymbol{\xi }_{i}^{A}(t)$ would then
be sampled from this initial nonequilibrium state of the environment and
consequently it would not represent standard thermal noise. This would also
necessarily imply that the initial state of the system would always be
dependent upon the realization of the noise. To obtain standard (initially
uncorrelated) noise from an initially equilibrium environment, the initial
state of the system must be of the form $\boldsymbol{\rho }_{\mathrm{sys}}(%
\mathbf{x},\mathbf{p})$, represented in terms of canonical coordinates $%
\mathbf{x}$ and $\mathbf{p.}$


\subsection{Electric-Dipole Gauge Hamiltonian}

\label{sec:QEDx}To obtain a Langevin equation with standard, statistically
independent noise, we can make the canonical transformation 
\begin{equation*}
e\,\dot{\mathbf{x}}\cdot \mathbf{A}(\mathbf{x})=e\,\mathbf{x}\cdot \dot{%
\mathbf{A}}(\mathbf{x})-e\frac{d}{dt}[\mathbf{x}\cdot \mathbf{A}(\mathbf{x}%
)]\,,
\end{equation*}%
and neglect the total derivative in the action (see, e.g. \cite%
{BaroneCaldeira91}). The Hamiltonian becomes 
\begin{align}
& \mathbf{H}_{\mathrm{QED}}^{\mbox{\normalsize$\pi$}}=\sum_{j}\left\{ \frac{%
\mathbf{p}_{j}^{2}}{2m_{j}}+e_{j}\,\phi (\mathbf{x}_{j})\right\} +
\label{eq:HQED2} \\
& \int \!d^{3}x^{\prime }\left\{ \frac{1}{2\varepsilon _{0}}[\boldsymbol{%
\mbox{\Large$\pi$}}(\mathbf{x}^{\prime })-\rho (\mathbf{x},\mathbf{x}%
^{\prime })\mathbf{x}]^{2}+\frac{\varepsilon _{0}c^{2}}{2}\left[ \boldsymbol{%
\nabla }\times \mathbf{A}(\mathbf{x}^{\prime })\right] ^{2}\right\} ,  \notag
\end{align}%
which is analogous to the QBM Hamiltonian \eqref{eq:HQBM}. The interaction
is 
\begin{equation*}
\mathbf{H}_{\mathrm{int}}=-\sum_{j}e_{j}\,\mathbf{x}_{j}\cdot \boldsymbol{%
\mbox{\Large$\pi$}}(\mathbf{x}_{j})\,,
\end{equation*}%
after the $\rho ^{2}\,\mathbf{x}^{2}$ term is included in the
renormalization of the potential $\phi$.

The equations of motion for the system driven by the field are 
\begin{align}
\dot{\mathbf{x}}_{j}& =\frac{\mathbf{p}_{j}}{m_{j}}\,,  \label{eq:Lx1} \\
\dot{\mathbf{p}}_{j}& =-e_{j}\boldsymbol{\nabla }_{\!\!j}\phi _{\mathrm{bare}%
}(\mathbf{x}_{j})+e_{j}\boldsymbol{\nabla }_{\!\!j}\left[ \mathbf{x}%
_{j}\cdot \boldsymbol{\mbox{\Large$\pi$}}(\mathbf{x}_{j})\right] \,,
\label{eq:Lx2}
\end{align}%
where we take the bare potential to include the divergent contribution from $%
\rho ^{2}\,\mathbf{x}^{2}$ in the Hamiltonian. The analogous analysis for
QBM is described in App.~\ref{sec:QBM}.

The equations of motion for the field driven by the particle are 
\begin{align}
\dot{\mathbf{A}}_{\mathbf{k},\boldsymbol{\epsilon }_{k}}& =\frac{1}{%
\varepsilon _{0}}\boldsymbol{\mbox{\Large$\pi$}}_{\mathbf{k},\boldsymbol{%
\epsilon }_{k}}-\sum_{j}e_{j}\cos (\mathbf{k}\cdot \mathbf{x}_{j})\,{\mathbf{%
x}}_{j}\,, \\
\dot{\boldsymbol{\mbox{\Large$\pi$}}}_{\mathbf{k},\boldsymbol{\epsilon }%
_{k}}& =-\varepsilon _{0}\omega _{k}^{2}\,\mathbf{A}_{\mathbf{k},\boldsymbol{%
\epsilon }_{k}}+\varepsilon _{0}\omega _{k}\sum_{j}e_{j}\sin (\mathbf{k}%
\cdot \mathbf{x}_{j})\,{\mathbf{x}}_{j}\,.
\end{align}%
As calculated in App.~\ref{sec:Explicit}, the driven solution 
expressed in the manifestly-Hermitian form is 
\begin{equation}
\boldsymbol{\mbox{\Large$\pi$}}(\mathbf{x}_{i},t)=\boldsymbol{\xi }_{i}^{%
\mbox{\normalsize$\pi$}}(t)-\sum_{j}e_{j}\left\{ (\boldsymbol{\mu }_{ij}^{%
\mbox{\normalsize$\pi$}}\ast \mathbf{x}_{j})(t)+(\boldsymbol{\mu }_{ij}^{%
\mbox{\normalsize$\pi$}}\ast \mathbf{x}_{j})^{\dagger }(t)\right\} ,
\label{eq:Pi(x)}
\end{equation}%
where the dissipation kernel $\boldsymbol{\mu }_{ij}^{\mbox{\normalsize$\pi$}%
}$ is resolved (to lowest orders in $1/c$) in Sec.~\ref{sec:gamma}.

In parallel to the analysis of the previous section, multiparticle Langevin
equations of motion can be obtained by substituting Eq.~\eqref{eq:Pi(x)} for 
$\boldsymbol{\mbox{\Large$\pi$}}$ into the system equations of motion %
\eqref{eq:Lx1}-\eqref{eq:Lx2}. Again, we delay writing them down because in
this approach the multiple-particle equations of motion are more cumbersome
that the form given in Section \ref{sec:EQED}. The single-particle case,
however, is analyzed in Sec.~\ref{sec:Standard}.

Unlike the analysis from the previous subsection, the field variables are
now expressed in terms of the canonical variable $\mathbf{x}_{j}$, in
contrast to $\dot{\mathbf{x}}_{i}$, and consequently the stochastic variable 
$\boldsymbol{\xi }_{i}^{\mbox{\normalsize$\pi$}}(t)$ can be interpreted as
statistically-independent noise\footnote{%
In fact, $\boldsymbol{\xi }_{i}^{\mbox{\normalsize$\pi$}}(t)$ can be
(perturbatively) identified with a stochastic electric field, though $%
\boldsymbol{\mbox{\Large$\pi$}}$ cannot be identified with the electric
field as it does not contain the electrostatic fields and necessarily
contains the magnetostatic fields.}. Issues due to noncommutativity of
system (particle) and environment (field) coordinates discussed in Sec.~\ref%
{sec:QEDv} do not occur in this gauge since an initial state of the form 
\begin{equation}
\boldsymbol{\rho }_{\mathrm{sys+env}}=\boldsymbol{\rho }_{\mathrm{sys}}(%
\mathbf{x},\mathbf{p})\otimes \boldsymbol{\rho }_{\mathrm{env}}(\mathbf{A},%
\boldsymbol{\mbox{\Large$\pi$}})
\end{equation}%
is consistent with the required initial data for particle position in the
Langevin equation, with $\boldsymbol{\rho }_{\mathrm{env}}(\mathbf{A},%
\boldsymbol{\mbox{\Large$\pi$}})$ at the same time an equilibrium state of
the field Hamiltonian. As will be reviewed in the next section, however, the
damping kernel for $\boldsymbol{\mu }_{ij}^{\mbox{\normalsize$\pi$}}$ is
more pathological than for $\boldsymbol{\mu }_{ij}^{A}$. The latter is
approximately a delta function (Ohmic), whereas the former is approximately
the second derivative of a delta function (supra-Ohmic).

To summarize, stochastic equations for motion obtained for the Coulomb gauge
with $\dot{\mathbf{x}}\cdot \mathbf{A}$ coupling in the particle-field
interaction have noise that is not statistically independent of the
particle's requisite initial data, and this will lead to severe
complications in both the interpretation and evaluation of the resulting
Langevin equations. In contrast, stochastic equations of motion for the
electric-dipole gauge with ${\mathbf{x}}\cdot \boldsymbol{\mbox{\Large$\pi$}}
$ coupling have statistically independent noise, but more pathological
damping, as described in the next section.


\section{Electromagnetic Damping Kernels}

\label{sec:gamma} We next analyze the nonlocal integration kernels which
arise for the Langevin equations. It is useful to first define the
commutators of vector operators 
\begin{equation*}
\lbrack \![\mathbf{X},\mathbf{Y}]\!]\equiv \mathbf{X}\,\mathbf{Y}^{\mathrm{T}%
}-\mathbf{Y}\,\mathbf{X}^{\mathrm{T}}\,.
\end{equation*}%
This object is a matrix whose entries are ordinary commutators. In this
notation, the dissipation kernel associated with $\mathbf{A}$-coupling from
Eq.~\eqref{eq:A(v)} is exactly specified in App.~\ref{sec:Explicit} and is
approximately given by the field commutator 
\begin{equation*}
\boldsymbol{\mu }_{\!A}[\mathbf{x}(t),\mathbf{y}(t^{\prime });t,t^{\prime
}]\approx \left\langle \frac{1}{2 \imath \hbar}\left[ \!\left[ \underline{%
\mathbf{A}}(\mathbf{x}(t),t),\underline{\mathbf{A}}(\mathbf{y}(t^{\prime
}),t^{\prime })\right] \!\right] \right\rangle _{\mathrm{\!\!env}} ,
\end{equation*}%
with phase discrepancies of the order $\mathcal{O}\left(
\hbar^2/c^{3}\right) $. These phase discrepancies denote
quantum-relativistic deviations from the quasistatic approximation wherein
we take $\mathbf{x}(t)$ and $\mathbf{y}(t^{\prime })$ to be adiabatic
variables inside the field correlations. This is largely equivalent to the
dipole approximation for one particle, but it allows us to systemically
extend the analysis to the multiparticle case while keeping tracking of the
order of the approximation in powers of $1/c$. More details on this
relativistic ``multipole'' expansion are given in App.~\ref{sec:Explicit}.

Let us denote the time dependence in the field operators which arises from $%
\mathbf{x}(t)$ as \emph{intrinsic} and the explicit time dependence in $%
\mathbf{A}(\mathbf{r},t)$ at some fixed location $\mathbf{r}$ as \emph{%
extrinsic}. With this distinction, the dissipation kernel is stationary with
regard to extrinsic time dependence, i.e., 
\begin{equation*}
\boldsymbol{\mu }_{\!A}[\mathbf{x}(t),\mathbf{y}(t^{\prime });t,t^{\prime }]=%
\boldsymbol{\mu }_{\!A}[\mathbf{x}(t),\mathbf{y}(t^{\prime
});t\!-\!t^{\prime }]\,.
\end{equation*}%
It is not stationary, however, with regard to the intrinsic time dependence
of $\mathbf{x}(t)$ and $\mathbf{y}(t^{\prime })$. The quantum dissipation
kernel, however, is approximately spatially stationary in the sense that 
\begin{equation*}
\boldsymbol{\mu }_{\!A}[\mathbf{x}(t),\mathbf{y}(t^{\prime });t,t^{\prime
}]\approx \boldsymbol{\mu }_{\!A}[\mathbf{x}(t)\!-\!\mathbf{y}(t^{\prime
});t\!-\!t^{\prime }]\,,
\end{equation*}%
with $\mathcal{O}\left( \hbar ^{2}/c^{3}\right) $ phase discrepancies as
discussed in App.~\ref{sec:Explicit}.

The positive-definite and Hermitian damping kernel is then given by 
\begin{align}
\boldsymbol{\mu }[\mathbf{x}(t),\mathbf{y}(t^{\prime });t,t^{\prime }]&
\equiv -\frac{\partial }{\partial t^{\prime }}\boldsymbol{\mu }[\mathbf{x}%
(t),\mathbf{y}(t^{\prime });t,t^{\prime }]\,, \\
\tilde{\boldsymbol{\mu }}[\mathbf{x}(t),\mathbf{y}(t^{\prime });\omega ]&
\equiv \imath \omega \,\tilde{\boldsymbol{\gamma }}[\mathbf{x}(t),\mathbf{y}%
(t^{\prime });\omega ]\,,
\end{align}%
where the partial derivative and Fourier transform neglect any intrinsic
time dependence in $\mathbf{x}(t)$ and $\mathbf{y}(t^{\prime })$, similar to
the analysis in \cite{Hsiang11}. As discussed in App.~\ref{sec:Explicit}, we
may neglect the time dependence intrinsic to $\mathbf{y}(t^{\prime })$ when
taking a total derivative, as the corrections are of higher order in $v/c$.

Fourier transforming with respect to the extrinsic time variables, the
classical or quasistatic $\mathbf{A}$-coupling damping kernel (without
cutoff) is 
\begin{align}
\tilde{\boldsymbol{\gamma }}_{\!A}[\mathbf{r};\omega ]& =2\gamma _{0}\left\{ 
\tilde{\mathrm{S}}_{1}\!\left( \left\vert \frac{\mathbf{r}\omega }{c}%
\right\vert \right) +\tilde{\mathrm{S}}_{0}\!\left( \left\vert \frac{\mathbf{%
r}\omega }{c}\right\vert \right) \,\hat{\mathbf{r}}\,\hat{\mathbf{r}}^{%
\mathrm{T}}\right\} \,,  \label{eq:EMgamma} \\
\gamma _{0}& =\frac{1}{12\pi \varepsilon _{0}c^{3}}\,,
\end{align}%
in terms of the functions 
\begin{align}
\tilde{\mathrm{S}}_{1}(z)& \equiv +\frac{3}{2}\frac{(z^{2}-1)\sin (z)+z\cos
(z)}{z^{3}}\,, \\
\tilde{\mathrm{S}}_{0}(z)& \equiv -\frac{3}{2}\frac{(z^{2}-3)\sin
(z)+3\,z\cos (z)}{z^{3}}\,.
\end{align}
A cutoff regulator $\chi$ can be easily inserted by multiplying the
right-hand-side of Eq.~\eqref{eq:EMgamma} with a function 
\begin{equation}
\chi(\omega/\Lambda) : [0,\infty) \mapsto [1,0) \, ,
\end{equation}
that vanishes sufficiently fast.

\begin{figure}
\includegraphics[width=0.5\textwidth]{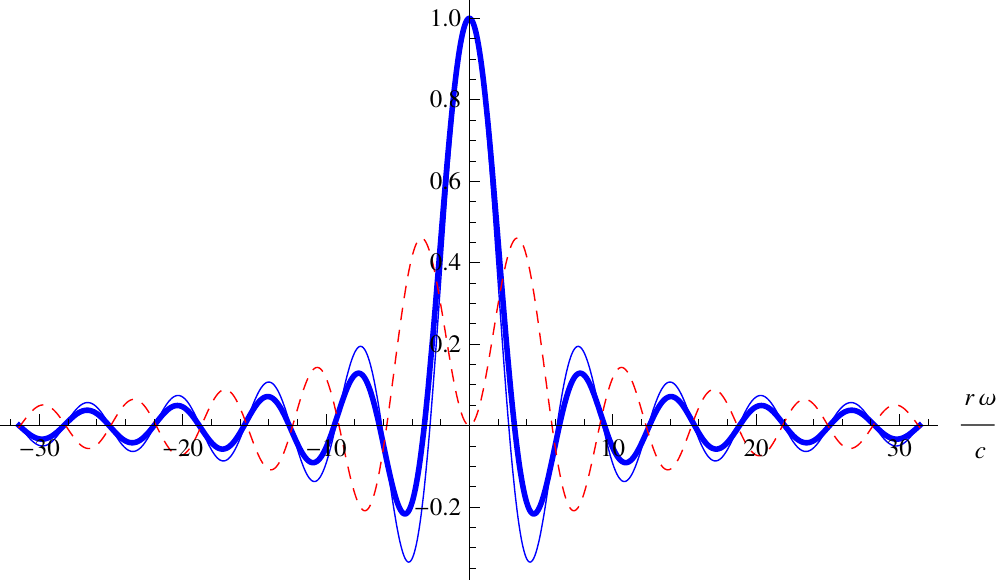} 
\caption{Comparison of sinc (bold), $\tilde{\mathrm{S}}_1$, and $\tilde{\mathrm{S}}_0$ (dashed).
Sinc and $\tilde{\mathrm{S}}_1$ are extremely qualitatively similar, both being unity at zero whereas $\mathrm{S}_0$ vanishes at zero.}
\label{fig:Polarity}
\end{figure}

\begin{figure}
\includegraphics[width=0.5\textwidth]{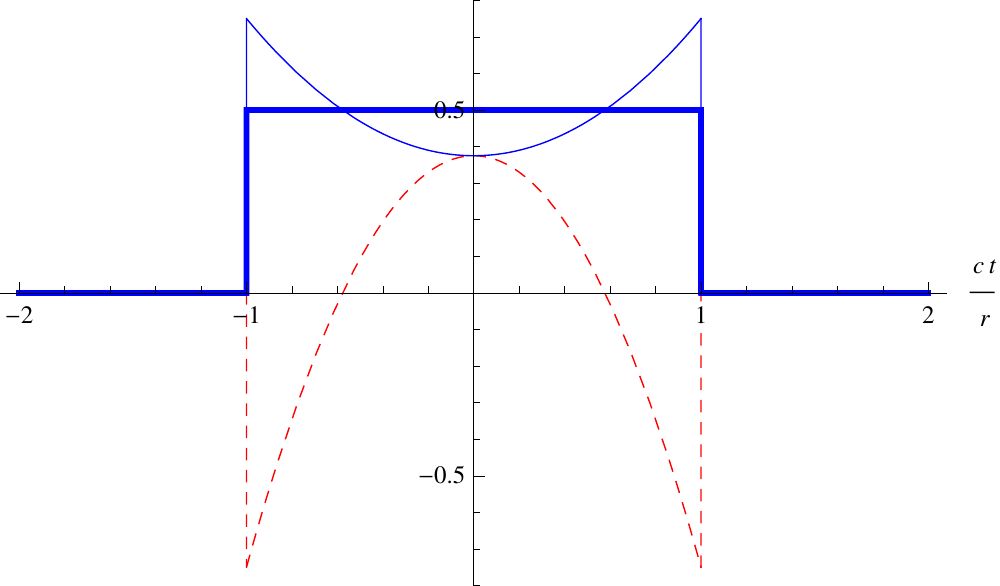} 
\caption{The same functions as
in Fig.~\ref{fig:Polarity}, but in the time domain: the rectilinear
distribution (bold), S${}_0$, and S${}_1$ (dashed).} \label{fig:Polarity2}
\end{figure}

In the coincidence limit we recover the usual Ohmic damping: 
\begin{align}
\lim_{r\rightarrow 0}\tilde{\boldsymbol{\gamma }}_{\!A}[\mathbf{r};\omega ]&
=2\gamma _{0}\,, \\
\lim_{r\rightarrow 0}{\boldsymbol{\gamma }}_{\!A}[\mathbf{r};t]& =2\gamma
_{0}\,\delta (t)\,,
\end{align}%
and for particles that are far separated all cross correlations vanish: 
\begin{align}
\lim_{r\rightarrow \infty }\tilde{\boldsymbol{\gamma }}_{\!A}[\mathbf{r}%
;\omega ]& =0\,, \\
\lim_{r\rightarrow \infty }{\boldsymbol{\gamma }}_{\!A}[\mathbf{r};t]& =0\,.
\end{align}%
In Fig.~\ref{fig:Polarity} we compare these special functions to $\mathrm{%
sinc}(z)$, which is the result obtained from the simpler but analogous case
of coupling to a scalar field. In Fig.~\ref{fig:Polarity2} we compare these
functions in the time domain. The frequency domain is useful for noting the
Markovian limit at $r\omega =0$ and the decorrelation which occurs for $%
r\omega \rightarrow \infty $. The time domain is useful for illustrating
that the damping kernel is (perturbatively) causal and constrained to the
light cone. As is well known, this latter property is not true of the
quantum noise kernel.

\subsection{Dissipation and Damping with $\mathbf{A}$-coupling}

\label{sec:magneto} To describe quasirelativistic damping, we can integrate
by parts the $\mathbf{A}$-coupling dissipation kernel to obtain terms
corresponding to damping, mass renormalization, and dissipationless
backreaction. In the quasirelativistic regime we treat the positions $%
\mathbf{x}$ as quasistatic within the convolution (as discussed in App.~\ref%
{sec:Explicit}) 
\begin{equation*}
(\boldsymbol{\mu }_{ij}^{A}\!\ast \mathbf{z})(t)=\int_{0}^{t}\!\!dt^{\prime }%
\boldsymbol{\mu }_{\!A}[\mathbf{x}_{i}(t)\!-\!\mathbf{x}_{j}(t^{\prime
});t\!-\!t^{\prime }]\,\mathbf{z}(t^{\prime })\,,
\end{equation*}%
where $\mathbf{z}(t)$ is some arbitrary source. Integrating by parts, and
neglecting the intrinsic time dependence in the positions (which is
equivalent to ignoring higher-order corrections in $v/c$), gives 
\begin{align*}
(\boldsymbol{\mu }_{ij}^{A}\!\ast \mathbf{z})(t)& =(\boldsymbol{\gamma }%
_{ij}^{A}\!\ast \dot{\mathbf{z}})(t)-\boldsymbol{\gamma }_{\!A}[\mathbf{x}%
_{i}(t)\!-\!\mathbf{x}_{j}(t);0]\,\mathbf{z}(t) \\
& +\boldsymbol{\gamma }_{\!A}[\mathbf{x}_{i}(t)\!-\!\mathbf{x}_{j}(0);t]\,%
\mathbf{z}(0)\,,
\end{align*}%
where the final term is a transient \emph{slip}. Although the slip evolution
arises during the same period of time as the conventional Abraham-Lorentz
acausality, its role is well understood from QBM studies where it has been
shown to be unproblematic (see App.~\ref{sec:QBM} and Ref.~\cite{QBM} where
the initial evolution for initial factorized system plus environment states
in terms of so-called \textquotedblleft slip\textquotedblright\ and
\textquotedblleft jolt\textquotedblright\ evolution is reviewed). It may
therefore be discarded for our purposes; doing so is equivalent to choosing
an initial state which is more properly correlated to the environment. The
difference between QBM and QED is that with the relativistic field, not only
are there instantaneous self-transients from $\boldsymbol{\gamma }_{ii}(t)$,
there are also retarded cross-transients from $\boldsymbol{\gamma }_{ij}(t)$%
. Figure~\ref{fig:Polarity2} shows that the electrodynamic slip only
produces a transient effect for $t<r/c$. Essentially, for factorized states
of the particles and field, the particles are completely unaware of each
other's existence until photons travel between them and suddenly (and
violently) establish correlations at their first mediation.

Unlike QBM, $\boldsymbol{\gamma }_{A}[\mathbf{r}_{ij}(t);0]$ contains both
self interactions, which renormalize the mass, and field-mediated
interactions. Using Eq.~\eqref{eq:EMgamma}, this term becomes 
\begin{equation*}
\boldsymbol{\gamma }_{\!A}[\mathbf{r};0]=\frac{3}{4}c\,\gamma _{0}\frac{1+%
\hat{\mathbf{r}}\,\hat{\mathbf{r}}^{\mathrm{T}}}{|\mathbf{r}|}\,,
\end{equation*}%
and we see the emergence of the Darwin Hamiltonian (e.g. see Jackson \cite%
{Jackson98} Sec. 12.7). The magnetostatic energy of the Darwin Hamiltonian
(which is classical) is given by 
\begin{equation*}
\mathbf{V}_{ij}=-2\frac{e_{i}}{m_{j}}\mathbf{p}_{i}^{\mathrm{T}}\boldsymbol{%
\gamma }_{\!A}[\mathbf{r}_{ij};0]\frac{e_{j}}{m_{j}}\mathbf{p}_{j}\,.
\end{equation*}%
The quantum extension of the Darwin Hamiltonian requires an
operator-ordering prescription that our analysis will provide upon applying
these results to the Langevin equation.

These results exemplify how the equations of motion that result from an $%
\mathbf{A}$-coupling interaction give consistent formulas for damping and
backreaction, despite the noise being problematic as previously discussed.


\subsection{Dissipation and Damping with Electric-Dipole Coupling}

In this subsection we describe the sense in which the electric-dipole
interaction gives a less desirable description of damping and backreaction,
despite providing a consistent description of thermal noise, as previously
described. For $\boldsymbol{\mbox{\Large$\pi$}}$-coupling to the field, the
dissipation kernel is approximately given by 
\begin{equation*}
\boldsymbol{\mu }_{\mbox{\normalsize$\pi$}}[\mathbf{x}(t),\mathbf{y}%
(t^{\prime });t,t^{\prime }]\approx \left\langle \frac{1}{2\imath \hbar }%
\left[ \!\left[ \underline{\boldsymbol{\mbox{\Large$\pi$}}}(\mathbf{x}(t),t),%
\underline{\boldsymbol{\mbox{\Large$\pi$}}}(\mathbf{y}(t^{\prime
}),t^{\prime })\right] \!\right] \right\rangle _{\mathrm{\!\!env}},
\end{equation*}%
with the same $\mathcal{O}\left( \hbar ^{2}/c^{3}\right) $ phase
discrepancies discussed in App.~\ref{sec:Explicit}. The $\boldsymbol{%
\mbox{\Large$\pi$}}$-coupling dissipation kernel is related to the $\mathbf{A%
}$-coupling dissipation kernel via 
\begin{align}
{\boldsymbol{\gamma }}_{\mbox{\normalsize$\pi$}}[\mathbf{r};t,t^{\prime }]&
=-\frac{\partial ^{2}}{\partial t^{\prime 2}}\,{\boldsymbol{\gamma }}_{\!A}[%
\mathbf{r};t,t^{\prime }]\,, \\
\tilde{\boldsymbol{\gamma }}_{\mbox{\normalsize$\pi$}}[\mathbf{r};\omega ]&
=\omega ^{2}\,\tilde{\boldsymbol{\gamma }}_{\!A}[\mathbf{r};\omega ]\,.
\end{align}%
This result is equivalent to dissipation from a supra-Ohmic bath in QBM. As
an integration kernel ${\boldsymbol{\gamma }}_{\mbox{\normalsize$\pi$}}$ is
relatively pathological and must be integrated by parts twice to obtain the
well-behaved kernel ${\boldsymbol{\gamma }}_{A}$. Such an integration is
straightforward for the single-particle theory in the dipole approximation,
but for the multiparticle and higher-order relativistic theory this
generates many additional terms and limits which must be carefully analyzed
in constructing the Langevin equation. Although, with care, it should be
possible to proceed this way, the approach in the next section is clearer
and more straightforward.


\subsection{Standard Dipole Calculation}

\label{sec:Standard} Before presenting our effective equations of motion in Sec.~\ref{sec:EQED}, we first review
the derivation of the standard results for a single particle. Taking the
dipole approximation we drop all position dependence in the integration
kernels. 
Substituting Eq.~\eqref{eq:Pi(x)} (integrated by parts once) into Eqs. %
\eqref{eq:Lx1}-\eqref{eq:Lx2} we obtain the Langevin equation 
\begin{equation}
m\ddot{\mathbf{x}}(t)=-e\boldsymbol{\nabla }\phi (\mathbf{x})-2\,e^{2}\,({%
\gamma }^{\mbox{\normalsize$\pi$}}\ast \dot{\mathbf{x}})(t)+e\,\boldsymbol{%
\xi }^{\mbox{\normalsize$\pi$}}(t)\,,  \label{eq:SL1}
\end{equation}%
discarding the transient terms. This equation of motion is runaway free for
bare mass $m>0$ as reviewed in App.~\ref{sec:QBMstable}. 

Integrating-by-parts the $\pi $-damping integral in the Langevin equation %
\eqref{eq:SL1} two times, to obtain $A$-coupling damping, discarding
additional transient terms, and grouping like terms, we obtain the standard
Abraham-Lorentz-Langevin equation 
\begin{align}
m_{\mathrm{ren}}\ddot{\mathbf{x}}(t)& =-e\boldsymbol{\nabla }\phi (\mathbf{x}%
)+2\,e^{2}(\gamma ^{A}\ast \dddot{\mathbf{x}})(t)-e\,\dot{\boldsymbol{\xi }}%
^{A}(t)\,,  \label{eq:LNR} \\
m_{\mathrm{ren}}& =m+2\,e^{2}\gamma ^{A}(0),  \label{eq:mass}
\end{align}%
where $m_{\text{ren}}$ is the renormalized particle mass. The same result is
obtained by integrating-by-parts the dissipation integral in the Langevin
equations \eqref{eq:Lv1}-\eqref{eq:Lv2}. 
Note that $(\gamma ^{A}\ast \dddot{\mathbf{x}})(t)\approx \gamma _{0}\,%
\dddot{\mathbf{x}}(t)$ in the high cutoff (\textquotedblleft
point-particle\textquotedblright ) limit. Positive bare mass $m,$ which is
required for runaway-free motion, requires a finite cutoff $\Lambda $ in the
field modes such that $2\,e^{2}\gamma ^{A}(0)<m_{\mathrm{ren}}\,,$ noting
that $\gamma ^{A}(0)\propto \gamma _{0}\,\Lambda $. Therefore positive bare
mass requires an ultraviolet cutoff $\Lambda ,$ or equivalently a form
factor that cuts off the particle field coupling, on the order of $%
\varepsilon _{0}c^{3}m_{\mathrm{ren}}/e^{2}$ or smaller. This is directly
seen from our stability analysis in App.~\ref{sec:QBMstable}, or simply by
noting that with a negative bare mass the system + environment Hamiltonian
no longer has a lower bound in its energy spectrum.


\section{$1/c$ Expansion of QED}

\label{sec:EQED}

In the previous section, we contrasted how the standard results are obtained
when the interaction is expressed in terms of either $\dot{\mathbf{x}}\cdot 
\mathbf{A}$ or $\mathbf{x}\cdot \boldsymbol{\mbox{\Large$\pi$}}$ coupling.
We have highlighted the problem of simultaneously obtaining well-behaved
noise and backreaction. In \cite{Ford91}, Ford and O'Connell showed that the
equations of motion for a charged particle with structure can be
order-reduced to obtain the structure-independent equation of motion 
\begin{equation}
m\ddot{\mathbf{x}}=\left\{ 1+\frac{2\,e^{2}\gamma _{0}}{m}\frac{d}{dt}%
\right\} \left[ \mathbf{F}_{\mathrm{ext}}(\mathbf{x})-e\,\dot{\boldsymbol{%
\xi }}(t)\right] \,,  \label{eq:FO}
\end{equation}%
which is accurate to lowest order in the timescale 
\begin{equation}
\tau _{m}=e^{2}/6\pi \varepsilon _{0}mc^{3}\sim \alpha \hbar /mc^{2}\,,
\end{equation}%
where $\alpha =e^{2}/4\pi \varepsilon _{0}\hbar c$ is the fine-structure
constant. For bound states, this is equivalent to the dimensionless order $%
\alpha \,\Delta E/mc^{2}$, where $\Delta E$ denotes the relevant
energy-level transitions of the system. For driving forces, the expansion
parameter is $\alpha \,\hbar \omega /mc^{2}$, given a driving frequency $%
\omega $. In either case the result is of order $1/c^{3}$, and the $e^{3}%
\dot{\boldsymbol{\xi }}$ term is negligible given the approximations which
have been made in constructing this equation.

Eq.~\eqref{eq:FO} is causal, runaway-free, and is in general dissipative
(i.e. accelerated motion is damped). The Ford and O'Connell result is
generic in the sense that any particle, with reasonable assumptions about
its structure (form factor), will have backreaction of this form in the
weak-backreaction limit. In this sense, the result is the universal
(effective) low-energy, and equivalently lowest order in $1/c$, result for
backreaction. Below, we obtain stochastic equations of motion with damping
consistent with Eq.~\eqref{eq:FO}, but from a different approach that
provides additional insights into the nature of backreaction. Whereas the
more-standard approach of Ford and O'Connell considers Eq.~\eqref{eq:ALL1}
to be of order $\mathcal{O}(1/c^{3})$ and their Eq.~\eqref{eq:FO} to be
order reduced, our approach will generate equivalent damping to Eq.~%
\eqref{eq:FO} which is strictly $\mathcal{O}(1/c^{3})$, and not order
reduced. Within our analysis, the standard Eq.~\eqref{eq:ALL1} is
mixed-ordered, and this is the primary source of its pathological behaviors.
An advantage of our method is that it allows for a consistent and systematic
extension to higher orders which includes both higher-order relativistic
corrections, and higher-order effects of backreaction.

Our approach fundamentally relies upon consistent application of a
dimensionful $1/c$ 
expansion around the Hamiltonian 
\begin{equation*}
\mathbf{H}_{\text{NR}}=\sum_{i}\frac{\mathbf{p}_{i}^{2}}{2m_{i}}+\sum_{i<j}%
\frac{1}{4\pi \varepsilon _{0}}\frac{e_{i}e_{j}}{\left\vert \mathbf{r}%
_{ij}\right\vert }\,,
\end{equation*}%
which is viewed as the lowest-order, \textquotedblleft
nonrelativistic\textquotedblright\ Hamiltonian in this approach. All of the
additional terms that arise in the QED Hamiltonian are then viewed as
perturbations, whose order is based on the powers of $1/c$.

%
We begin with the $\mathbf{A}$-coupling, Coulomb-gauge Hamiltonian in %
\eqref{eq:HQED1}. 
From the previous analysis we observe that the self-field generates, at
lowest order $\mathcal{O}(1/c^{2}),$ magnetostatic and renormalization terms
(see Sec.~\ref{sec:magneto}). Because our approximation is based on a
systematic expansion in powers of $1/c,$ and here we work to $\mathcal{O}
(1/c^{3}),$ we effectively drop the $\mathbf{A}^{2}$ system-field
interaction terms in the Hamiltonian.\footnote{%
For external fields this approximation should not be made, and any $\mathbf{A%
}_\mathrm{ext}$ should be included by appropriately translating the system
momenta.} These neglected terms are all of order $(v/c)^{4}$, $\alpha
\,(\Delta E/mc^{2})\,(v/c)^{2}$, and so, making them $\mathcal{O}(1/c^{4})$.
For now we assume the usual nonrelativistic kinetic energy for the
particles, but for consistency we include the relativistic corrections at
the appropriate order in Sec.~\ref{sec:kinematic}. Therefore, the consistent 
$\mathcal{O} (1/c^{3})$ \emph{effective} Hamiltonian in the Coulomb gauge is 
\begin{align}
& \mathbf{H}_{\mathrm{EQED}}=\sum_{i}\left\{ \frac{\mathbf{p}_{i}^{2}}{2m_{i}%
}+e\,\phi (\mathbf{x}_{i})\right\}  \label{eq:HEQED} \\
& -\sum_{i}\frac{e_{i}}{2m_{i}}\left\{ \mathbf{p}_{i}\cdot \mathbf{A}(%
\mathbf{x}_{i})+\mathbf{A}(\mathbf{x}_{i})\cdot \mathbf{p}_{i}\right\} + 
\mathbf{H}_{\mathrm{field}}+\mathcal{O}(1/c^{4})\,.  \notag
\end{align}

The Heisenberg equations of motion for the system driven by the field for
the effective Hamiltonian $\mathbf{H}_{\mathrm{EQED}}$ are 
\begin{align}
\dot{\mathbf{x}}_{i}& =\frac{\mathbf{p}_{i}}{m_{i}}-\frac{e_{i}}{m_{i}}%
\mathbf{A}(\mathbf{x}_{i})\,,  \label{eq:Lv1N} \\
\dot{\mathbf{p}}_{i}& =-e_{i}\boldsymbol{\nabla }_{\!\!i}\phi (\mathbf{x}%
_{i})+\frac{e_{i}}{2m_{i}}\boldsymbol{\nabla }_{\!\!i}\left\{ \mathbf{p}%
_{i}\cdot \mathbf{A}(\mathbf{x}_{i})+\mathbf{A}(\mathbf{x}_{i})\cdot \mathbf{%
p}_{i}\right\} \,,  \label{eq:Lv2N}
\end{align}%
and the Heisenberg equations of motion for the environment driven by the
system are 
\begin{align}
\dot{\mathbf{A}}_{\mathbf{k},\boldsymbol{\epsilon }_{k}}& =\frac{1}{%
\varepsilon _{0}}\boldsymbol{\mbox{\Large$\pi$}}_{\mathbf{k},\boldsymbol{%
\epsilon }_{k}}+\frac{1}{\varepsilon _{0}\omega _{k}}\sum_{j}\frac{e_{j}}{%
2m_{j}}\left\{ \sin (\mathbf{k}\cdot \mathbf{x}_{j}),{\mathbf{p}}%
_{j}\right\} \,, \\
\dot{\boldsymbol{\mbox{\Large$\pi$}}}_{\mathbf{k},\boldsymbol{\epsilon }%
_{k}}& =-\varepsilon _{0}\omega _{k}^{2}\,\mathbf{A}_{\mathbf{k},\boldsymbol{%
\epsilon }_{k}}+\sum_{j}\frac{e_{j}}{2m_{j}}\left\{ \cos (\mathbf{k}\cdot 
\mathbf{x}_{j}),{\mathbf{p}}_{j}\right\} \,.
\end{align}%
As derived in App.~\ref{sec:Explicit}, the driven solution expressed in
manifestly-Hermitian form is 
\begin{equation}
\mathbf{A}(\mathbf{x}_{i},t)=\boldsymbol{\xi }_{i}^{A}(t)-\sum_{j}\frac{e_{j}%
}{m_{j}}\left\{ (\boldsymbol{\mu }_{ij}^{A}\ast {\mathbf{p}}_{j})(t)+(%
\boldsymbol{\mu }_{ij}^{A}\ast {\mathbf{p}}_{j})^{\dagger }(t)\right\} \,,
\label{eq:A(p)}
\end{equation}%
where the $\boldsymbol{\xi }_{i}(t)$ are uncorrelated processes for $i\neq j$%
. Following the approach in Sec.~\ref{sec:magneto}, integrating the
dissipation kernel by parts gives 
\begin{align}
& \mathbf{A}[\mathbf{x}_{i}(t),t]=\boldsymbol{\xi }_{i}(t)-2\frac{e_{i}}{%
m_{i}}\gamma _{0}\,\dot{\mathbf{p}}_{i}(t)  \label{eq:A(p)2} \\
& +\sum_{j}\left\{ \boldsymbol{\gamma }[\mathbf{r}_{ij}(t);0]\frac{e_{j}}{%
m_{j}}\mathbf{p}_{j}(t)+\left( \boldsymbol{\gamma }[\mathbf{r}_{ij}(t);0]%
\frac{e_{j}}{m_{j}}\mathbf{p}_{j}(t)\right) ^{\!\dagger }\right\} ,  \notag
\end{align}%
Here we have neglected the transient slip terms (which, as noted before and
as is reviewed in the appendix, can be verified to not induce runaways or
acausal behavior) and corrections of higher order in $v/c$.

Substituting the driven field solution into the system equations of motion
gives the renormalized multiple-particle Langevin equation 
\begin{align}
\dot{\mathbf{x}}_{i}=\;& \frac{\mathbf{p}_{i}}{m_{i}}-\frac{e_{i}}{m_{i}}%
\boldsymbol{\xi }_{i}+2\frac{e_{i}^{2}}{m_{i}^{2}}\gamma _{0}\,\dot{\mathbf{p%
}}_{i}+\boldsymbol{\nabla }_{\!\!\mathbf{p}_{i}}\sum_{j\neq i}\mathbf{V}%
_{ij}\,,  \label{eq:Lfinal1} \\
\dot{\mathbf{p}}_{i}=\;& -e_{i}\boldsymbol{\nabla }_{\!\!\mathbf{x}_{i}}\phi
(\mathbf{x}_{i})-\boldsymbol{\nabla }_{\!\!\mathbf{x}_{i}}\sum_{j\neq i}%
\mathbf{V}_{ij}\,,  \label{eq:Lfinal2}
\end{align}%
Details of the renormalization will be explained further in Sec.~\ref%
{sec:MREN}. The quantum magnetostatic potential $\mathbf{V}$ is given by 
\begin{align}
\mathbf{V}_{ij}& =-\frac{1}{2}\frac{e_{i}}{m_{i}}\frac{e_{j}}{m_{j}}\left[ 
\mathbf{p}_{i}^{\mathrm{T}}\boldsymbol{\gamma }[\mathbf{r}_{ij};0]\,\mathbf{p%
}_{j}+\mathbf{p}_{j}^{\mathrm{T}}\boldsymbol{\gamma }[\mathbf{r}_{ji};0]\,%
\mathbf{p}_{i}\right]  \notag \\
& -\frac{1}{2}\frac{e_{i}}{m_{i}}\frac{e_{j}}{m_{j}}\mathrm{Tr}_{x}\!\left[ 
\mathbf{p}_{i}\,\mathbf{p}_{j}^{\mathrm{T}}\,\boldsymbol{\gamma }[\mathbf{r}%
_{ij};0]+\boldsymbol{\gamma }[\mathbf{r}_{ij};0]\,\mathbf{p}_{i}\,\mathbf{p}%
_{j}^{\mathrm{T}}\right] \,.  \label{eq:VOO}
\end{align}%
The last two terms have been expressed with a spatial trace 
\begin{equation}
\mathrm{Tr}_{x}[\mathbf{M}] \equiv M_{xx} + M_{yy} + M_{zz} \, ,
\end{equation}
to keep the proper Hilbert-space operator ordering. The complexity of this
expression is due to the non-commutativity from both Hilbert-space and
spatial (3-vector and 3$\times $3 matrix) operations. This quantum
magnetostatic potential is consistent with the (classical) Darwin
magnetostatic potential (also see Sec.~\ref{sec:magneto}) 
\begin{align}
\mathbf{V}_{ij}& =-2\frac{e_{i}}{m_{i}}\mathbf{p}_{i}^{\mathrm{T}}%
\boldsymbol{\gamma }[\mathbf{r}_{ij};0]\frac{e_{j}}{m_{j}}\mathbf{p}_{j}\,,
\label{eq:Magneto1} \\
\boldsymbol{\gamma }[\mathbf{r};0]& =\frac{3}{4}c\,\gamma _{0}\frac{1+\hat{%
\mathbf{r}}\,\hat{\mathbf{r}}^{\mathrm{T}}}{r}\,.  \label{eq:Magneto2}
\end{align}%
Moreover, the operator ordering in Eq.~\eqref{eq:VOO} is consistent with the
Breit equation for spin-$1/2$ particles \cite{Breit30} (often named
\textquotedblleft orbit-orbit interaction\textquotedblright\ in that
context), when one ignores all of the Pauli spin matrices.

Equations \eqref{eq:Lfinal1}-\eqref{eq:Lfinal2} are one of the main results
of this paper. They are multiparticle stochastic equations of motion
consistent through order $\mathcal{O}\!\left( 1/c^{3}\right) $ in the field
influences with, as we will show, stable (runaway-free) backreaction even as
the field cutoff goes to infinity. [$\mathcal{O}\!\left( 1/c^{3}\right) $
kinematics will be given in Sec.~\ref{sec:kinematic}.] Notice that our
results show that the mass renormalization associated with the dissipative
backreaction, at $\mathcal{O}\!\left( 1/c^{3}\right) ,$ is due to
magnetostatic self-interactions. Our effective Hamiltonian treatment also
correctly produces the second-order magnetostatic corrections without
extraneous fourth-order terms (as Breit accidentally included in his first
calculation \cite{Breit29,Breit30}). This result is not present in the
standard treatments.

Our analysis suggests that the pathologies of the standard Abraham-Lorentz
equations can be viewed, from the perspective of a $1/c$ effective theory
expansion, as a consequence of performing a mixed-order calculation. In
terms of a $1/c$ expansion starting from the Coulomb-gauge Hamiltonian,
there are $\mathcal{O}\!\left( 1/c^{4}\right) $ \emph{multipole}-expansion
terms in the $\mathbf{p\cdot A}\!\left( \mathbf{x}\right) $ interaction
terms that are of the same order as the lowest-order \emph{dipole} terms
present in the $\mathbf{A}\!\left( \mathbf{x}\right) ^{2}$ interaction. This
reveals that uniform application of the dipole approximation by itself to
the Coulomb gauge Hamiltonian is inconsistent with the $1/c$ expansion, in
the sense that it keeps some terms of $\mathcal{O}\!\left( 1/c^{4}\right)$,
but discards others. For a consistent $\mathcal{O}\!\left( 1/c^{3}\right) $
expansion, effectively the dipole approximation to the $\mathbf{p\cdot A}%
\!\left( \mathbf{x}\right) $ interaction is kept and the $\mathbf{A}\!\left(%
\mathbf{x}\right) ^{2}$ interaction is dropped entirely, as we have done.

Historically, the Abraham-Lorentz equation has most often not been viewed
perturbatively, but instead as an effort to obtain an exact, nonrelativistic
point-particle result. Our philosophy has been to instead look for a
consistent low-energy, effective theory from the beginning. Giving up any
claim to an exact solution, we gain new insight into how a perturbatively
consistent solution arises within a framework that can be systematically
extended to higher orders. An alternative approach is to apply order
reduction to the standard result, which can be iterated into the
Ford-O'Connell equation \eqref{eq:FO}, with additional backreaction
contributions of higher order in $1/c$ (and higher-order derivatives). One
then discards these higher-order terms in the name of order reduction.
Combining our relations \eqref{eq:Lfinal1} and \eqref{eq:Lfinal2} reproduces
the same damping as the Ford-O'Connell equation, however, without any
higher-order terms to be order reduced. Thus, in terms of a $1/c$ expansion,
the damping in the Ford-O'Connell equation is strictly $1/c^{3}$ (and thus
consistent), whereas the pathological backreaction in the Abraham-Lorentz
equation is of mixed order. There is an interesting parallel here to Breit's
original calculation \cite{Breit29}, where pathological equations resulted
when Breit accidentally included a fourth-order operator in a calculation
which was only accurate to second order.

\subsection{Noise and backreaction stability}

\label{sec:QEDstable}

As previously discussed in Section \ref{sec:QEDv} the \textquotedblleft
noise\textquotedblright\ for the standard ${\mathbf{v}}\cdot \mathbf{A}$
coupling Langevin equation 
cannot be independently sampled from a thermal distribution. The standard
calculation only gives well-behaved noise within the electric-dipole gauge.
In contrast, our analysis, to order $\mathcal{O}\!\left( 1/c^{3}\right) $
starting from the Coulomb gauge, gives well-behaved noise and dissipation.
Comparing the derivations in Sections \ref{sec:QEDv} and \ref{sec:EQED}, we
see that our noise and the standard electric-dipole gauge noise only differ
by contributions from the $\mathbf{A}\!\left( \mathbf{x}\right) ^{2}$
interaction of the order $1/c^{4}$ and higher, or more specifically $%
(v/c)^{4}$, $\alpha \,(\Delta E/mc^{2})\,(v/c)^{2}$, etc..

It is interesting to consider further why the noise in the standard
(mixed-order) Coulomb gauge calculation is problematic. Examining Eq.~%
\eqref{eq:A(v)}, we see that the backreaction (which depends on velocity ${%
\mathbf{v}}$ in this expression) contains some perturbative amount of the
\textquotedblleft $\mathbf{p}$-noise\textquotedblright , which is true
thermal noise, obtained in our 
calculation. Similarly, inspection also shows that the noise in the standard
(mixed-order) Coulomb gauge calculation contains some perturbative amount of
\textquotedblleft $\mathbf{p}$-backreaction\textquotedblright , which is the
resistive damping that accompanies thermal noise, obtained in our
calculation. In other words, the (non-thermal) noise in the standard
(mixed-order) Coulomb gauge calculation contains some backreaction, and the
backreaction contains some noise. Therefore, even in the classical 
limit of \eqref{eq:A(v)} we would have to enforce an artificial constraint
upon the $\mathbf{p}$-backreaction implying that the backreaction in the
Coulomb gauge includes a \emph{non-zero} average value of the $\mathbf{p}$%
-noise.

The physics may be clearer from another perspective. Recalling that $m%
\mathbf{v}=\mathbf{p}-e\mathbf{A}$ such that the velocity $\mathbf{v}$
depends on both the particle and field canonical coordinates, we see that
velocity driving entails that the field is driven by both the system
(particle) and itself. In other words, velocity driving of the field implies
a perturbative feedback loop in the environment dynamics. Essentially, the
field can self-generate stronger and stronger field excitations, albeit at
an order beyond which the theory is accurate, and this can result in
runways. In contrast, in our new equations of motion, Eq. \eqref{eq:A(p)},
the field is driven by the canonical momentum $\mathbf{p}$ rather than the
velocity $\mathbf{v}$, and such pathological processes do not occur.

Our next task is to demonstrate that the equations of motion are
nonperturbatively stable. When combined, relations \eqref{eq:Lfinal1} and %
\eqref{eq:Lfinal2} reproduce the same damping as the Ford-O'Connell equation %
\eqref{eq:FO}, which is already known to be stable, and the noises are
perturbatively consistent. We can gain additional insight, however, by
comparing our analysis to the analysis leading to \eqref{eq:FO}. In this
paper, we begin with an effective Hamiltonian, and work consistently to $%
\mathcal{O}\!\left( 1/c^{3}\right) .$ With regard to the self-force this is
essentially equivalent to taking both the dipole approximation and
neglecting the $\mathbf{A}^{2}$ term in the Hamiltonian. At $\mathcal{O}%
\!\left(1/c^{3}\right) $, we find that there is no constraint on the cutoff
and the backreaction is fully insensitive to particle structure. In
contrast, in \eqref{eq:FO} the dipole approximation is made to the full
Hamiltonian including the $\mathbf{A}^{2}$ term, and instead a
particle-structure form factor is assumed, which gives an effective cutoff
in the particle-field interaction. The equations of motion are then order
reduced, effectively to $\mathcal{O}\!\left( 1/c^{3}\right) $, and the
resulting structure-independent equations of motion are found to be runaway
free. The ultimate agreement between these approaches is consistent with the
fact that the contribution to backreaction from the $\mathbf{A}^{2}$ term in
the Hamiltonian is effectively negligible in the regime where the two
methods of approximation are valid.

Although it is already known that Eq.~\eqref{eq:FO} is runaway free and
causal, we now re-analyze these properties within the present framework to
see how this physics arises from with a purely effective theory framework.
We will show that the dynamics of the open system are dissipative and stable
in a manner analogous to the QBM analysis given in App.~\ref{sec:QEDstable}.
Consider a single particle and denote the system Hamiltonian 
\begin{equation*}
\mathbf{H}_{\mathrm{sys}}\equiv \frac{\mathbf{p}^{2}}{2m}+e\,\phi (\mathbf{x}%
)\,.
\end{equation*}%
An energy constraint can be obtained from either the Heisenberg equations of
motion for $\mathbf{H}_{\mathrm{sys}}(t)$ or by integrating the (classical)
Langevin equation \eqref{eq:E2Bren} along with the second \eqref{eq:E2Bren2}%
. Discarding the irrelevant transient terms (which can be shown to be
bounded and runaway-free) we obtain the relation 
\begin{equation*}
\mathbf{H}_{\mathrm{sys}}(t)=\mathbf{H}_{\mathrm{sys}}(0)+\mathbf{H}_{\gamma
}(t)+\mathbf{H}_{\xi }(t)\,,
\end{equation*}%
where%
\begin{equation}
\mathbf{H}_{\gamma }(t)=-2\frac{e^{2}}{m^{2}}\gamma
_{0}\int_{0}^{t}\!\!dt^{\prime }\,\dot{\mathbf{p}}\!\left( t^{\prime
}\right) ^{2}
\end{equation}%
is the energy lost to damping and 
\begin{equation}
\mathbf{H}_{\xi }(t)=\frac{e}{m}\int_{0}^{t}\!\!dt^{\prime }\frac{1}{2}%
\left\{ \boldsymbol{\xi }(t^{\prime })\cdot \dot{\mathbf{p}}(t^{\prime })+%
\dot{\mathbf{p}}(t^{\prime })\cdot \boldsymbol{\xi }(t^{\prime })\right\}
\end{equation}%
is the work done by the noise $\boldsymbol{\xi }(t^{\prime }).$ The
contribution from damping is manifestly a negative quantity. The noise is
random and may do positive or negative work, but the damping only removes
energy from the system (and delivers it to the environment and interaction).
At least at this order, the damping is strictly local and so energy is lost
to dissipation in a strictly uniform manner.

It is important to note that the system here is given by the canonical
variables $\left( \mathbf{x,p}\right) $, and $\mathbf{H}_{\mathrm{sys}}$
does not correspond to the mechanical energy of the particle, for the same
reason that $\mathbf{p}^{2}/2m$ is not the mechanical kinetic energy. From
Eq.~\eqref{eq:Lv1} the velocity and momentum differ by the vector potential,
which includes both backreaction and noise. If the system momentum $\mathbf{p%
}$ relaxes under dissipative motion, then so does $\dot{\mathbf{p}}$ and by
extension the backreaction $\gamma _{0}\,\dot{\mathbf{p}}$. Given noise, the
system velocity fluctuates around the average 
\begin{equation}
\left\langle m\,\mathbf{v}\right\rangle _{\boldsymbol{\xi }}=\left\langle 
\mathbf{p}+2\frac{e^{2}}{m}\gamma _{0}\,\dot{\mathbf{p}}\right\rangle _{\!\!%
\boldsymbol{\xi }}\,,
\end{equation}%
implying that the system velocity is damped. If no external forces are
applied, the canonical and mechanical momenta approach each other (on
average) in the late-time limit. 
%

\subsection{Mass Renormalization}

\label{sec:MREN}In our Langevin equations \eqref{eq:Lfinal1}-%
\eqref{eq:Lfinal2}, the $\mathbf{p}^{2}/2m$ mass renormalization takes the
form of magnetostatic self-interaction, which is ordinarily discarded in classical
electrodynamics. Here we examine the counter terms involved in the
renormalization and contrast them to the standard mass renormalization
previously discussed in Sec.~\ref{sec:Standard}.

Consider most simply the single particle theory in the dipole approximation.
The resultant open-system equations of motion are then given by 
\begin{align}
\dot{\mathbf{x}}(t)& =\frac{\mathbf{p}(t)}{m_{\mathrm{ren}}}+2\frac{e^{2}}{%
m_{\mathrm{bare}}^{2}}\gamma _{0}\,\dot{\mathbf{p}}(t)-\frac{e}{m_{\mathrm{%
bare}}}\,\boldsymbol{\xi }(t)\,, \\
\dot{\mathbf{p}}(t)& =-e\boldsymbol{\nabla }\phi (\mathbf{x})\,,
\end{align}%
in terms of the renormalized mass 
\begin{equation}
\frac{1}{m_{\mathrm{ren}}}=\frac{1}{m_{\mathrm{bare}}}+2\frac{e^{2}}{m_{%
\mathrm{bare}}^{2}}\gamma (0)\,.  \label{eq:mren2}
\end{equation}%
Consistent with this order of perturbative analysis, we may express our
Langevin equations as 
\begin{align}
\dot{\mathbf{x}}(t)& =\frac{\mathbf{p}(t)}{m_{\mathrm{ren}}}+2\frac{e^{2}}{%
m_{\mathrm{ren}}^{2}}\gamma _{0}\,\dot{\mathbf{p}}(t)-\frac{e}{m_{\mathrm{ren%
}}}\,\boldsymbol{\xi }(t)\,,  \label{eq:E2Bren} \\
\dot{\mathbf{p}}(t)& =-e\boldsymbol{\nabla }\phi (\mathbf{x})\,.
\label{eq:E2Bren2}
\end{align}%
In this perspective, the renormalization of the $\mathbf{p}^{2}/2m$ mass and
the $e/m\,(\mathbf{p}\cdot \mathbf{A})$ mass enter at different orders. 
%

It is well known that the instability of the Abraham-Lorentz equation arises
from its implied negative bare mass for the system, which in-turn comes
about if the high frequency cutoff $\Lambda $ (or reciprocal radius $c/r$)
exceeds the characteristic frequency $\tau _{m}^{-1}$ \cite{Ford91,Moniz77}.
Yet to lowest order in $\tau _{m}$ (equivalently $1/c^{3}$), the dynamics of
charged-particle motion is insensitive to the high-energy details of the
theory and is thus not problematic in that regime \cite{Ford91}. It is
therefore not surprising that our perturbative approach yields
cutoff-insensitive behavior, however, the manner in which cutoff sensitivity
is avoided is interesting. Whereas the standard radiation-reaction
calculations involve a mass renormalization of $m\mathbf{v}^{2}/2$ given by
Eq.~\eqref{eq:mren2}, which runs the bare mass to negative infinity in the
high cutoff limit, our calculation runs the bare mass to positive zero in
the high cutoff limit, and no pathological behavior is induced for any
finite cutoff.


\subsection{Relativistic Kinematics}

\label{sec:kinematic} For consistency in our quasirelativistic expansion of $%
1/c$, we should include the relativistic corrections to the single particle
kinetic energy, as is standard in the Darwin Hamiltonian. Expanding the
relativistic kinetic energy gives 
\begin{align}
\mathbf{H}_{\mathrm{rel}}& =c\,\sqrt{(mc)^{2}+\left( \mathbf{p}-e\,\mathbf{A}%
\right) ^{2}}\,, \\
& \approx mc^{2}+\frac{\left( \mathbf{p}-e\,\mathbf{A}\right) ^{2}\!\!}{2m}-%
\frac{\left( \mathbf{p}-e\,\mathbf{A}\right) ^{4}\!\!}{8m^{3}c^{2}}+\cdots
\,,
\end{align}%
keeping terms of order $\mathcal{O}(1/c^{2})$. In the absence of external
fields, this generates a second-order correction to Eq.~\eqref{eq:Lfinal1}
giving 
\begin{equation}
\dot{\mathbf{x}}(t)=\left[ 1-\frac{1}{2}\left( \frac{\mathbf{p}(t)}{mc}%
\right) ^{2}\right] \frac{\mathbf{p}(t)}{m}+2\frac{e^{2}}{m^{2}}\gamma _{0}\,%
\dot{\mathbf{p}}(t)-\frac{e}{m}\,\boldsymbol{\xi }(t)\,,  \label{eq:E2Bren3}
\end{equation}%
and otherwise the Langevin equations are identical. As in the Darwin
Hamiltonian, this relativistic correction must be considered perturbatively.
At the present order of perturbation theory, however, we can resum the term
into the free velocity 
\begin{equation*}
\left[ 1-\frac{1}{2}\left( \frac{\mathbf{p}}{mc}\right) ^{2}\right] \frac{%
\mathbf{p}}{m}=\frac{\frac{\mathbf{p}}{m}}{\sqrt{1+\left( \frac{\mathbf{p}}{%
mc}\right) ^{2}}}+\mathcal{O}\!\left( \frac{1}{c^{4}}\right) \,,
\end{equation*}%
which is more convenient and better behaved in the equations of motion. This
can also be done in the effective Hamiltonian to ensure that the energy
spectrum has a lower bound.

Note that even nonperturbatively these kinematic corrections included in the
stability analysis given in Sec.~\ref{sec:QEDstable} will still yield
dissipative backreaction, as all of these terms commute with the $\mathbf{A}%
\cdot \mathbf{p}$ interaction at the relevant order. The only modification
will be in the definition of the canonical system Hamiltonian (now
relativistic), and the noise average of the velocity will be given by 
\begin{equation}
\left\langle m\,\mathbf{v}\right\rangle _{\boldsymbol{\xi }}=\left\langle 
\frac{\mathbf{p}}{\sqrt{1+\left( \frac{\mathbf{p}}{mc}\right) ^{2}}}+2\frac{%
e^{2}}{m}\gamma _{0}\,\dot{\mathbf{p}}\right\rangle _{\!\!\!\!\boldsymbol{%
\xi }}\,.
\end{equation}


\subsection{Electromagnetic Damping is Relativistic}

\label{sec:DampRel} The standard Abraham-Lorentz equation \eqref{eq:ALL1} is
commonly referred to as \textquotedblleft nonrelativistic\textquotedblright
. Within the framework of a $1/c$ expansion, this equation is more
accurately described as quasirelativistic. From this perspective, the
damping force, which is $\mathcal{O}(1/c^{3}),$ is intrinsically a
relativistic correction to the particle dynamics. 
%
Let us reexpress the $(v/c)^{2}$ magnetostatic and $1/c^{3}$ damping forces
in Eq.~\eqref{eq:Lfinal1} as both arising from dynamical generators 
\begin{align}
\dot{\mathbf{x}}_{i}=\;& \frac{\mathbf{p}_{i}}{m_{i}}-\frac{e_{i}}{m_{i}}%
\boldsymbol{\xi }_{i}+\boldsymbol{\nabla }_{\!\!\mathbf{p}_{i}}\left\{ 
\boldsymbol{\Gamma }_{i}+\sum_{j\neq i}\mathbf{V}_{ij}\right\} , \\
\boldsymbol{\Gamma }_{i}& \equiv \frac{d}{dt}\left( \frac{e_{i}}{m_{i}}%
\mathbf{p}_{i}^{\mathrm{T}}\gamma _{0}\frac{e_{i}}{m_{i}}\mathbf{p}%
_{i}\right) \,.
\end{align}%
By comparison with Eq.~\eqref{eq:Magneto1}-\eqref{eq:Magneto2}, if the
magnetostatic potential $\mathbf{V}$ is considered $(v/c)^{2}$, then the
damping generator $\boldsymbol{\Gamma }$ is of relative order $%
(r/c)(d/dt)(v/c)^{2}$. Thus the damping force can be interpreted as a more
dynamical and more relativistic correction. Given that both of these
generators arise from the same integration kernel without the full dipole
approximation (see App.~\ref{sec:Explicit}), a higher-order
\textquotedblleft multipole\textquotedblright\ expansion in $1/c$ can expect
to include many more such terms of higher order. The relativistic nature of
the expansion terms will be even clearer at higher orders.


\section{Summary and Discussions}

\label{sec:Discussion}

We have derived new stochastic equations of motion \eqref{eq:Lfinal1}-%
\eqref{eq:Lfinal2} \& \eqref{eq:E2Bren3} for multiple charged particles in
the electromagnetic field. These equations of motion incorporate the known
relativistic corrections to the electrodynamics of spinless charged
particles to order $\mathcal{O}(1/c^{3})$, including the electrostatic,
magnetostatic, electromagnetic damping forces, and field fluctuations.
Moreover the equations of motion are manifestly causal and runaway-free. Our
analysis shows that a $1/c$ expansion to the Coulomb-gauge Hamiltonian
describes consistent and well-behaved nonequilibrium electrodynamics for
spinless charged particles. Our consistent-order equations of motion have a
close correspondence with the order-reduced Ford-O'Connell equation, which
is known to be well behaved. Whereas traditionally the Abraham-Lorentz
equation has been considered an \textquotedblleft exact
nonrelativistic\textquotedblright\ equation, from the perspective of our
analysis, pathologies in the standard Abraham-Lorentz equation are
associated with inconsistent, mixed-order in $1/c$, approximations. Our view
is that radiation reaction is intrinsically relativistic, though the
Abraham-Lorentz equation may only fully capture its effect to lowest order
in $1/c$, and the order reduction used in deriving the Ford-O'Connell
equation actually serves to rid the dynamics of inappropriate mixed-order
contributions. 

At $\mathcal{O}\!\left( 1/c^{3}\right)$, mass renormalization is identified
with the magnetostatic self-interaction and, at this order, the bare mass is
positive for all cutoffs and the backreaction is fully insensitive to
particle structure. An interesting question for future research is whether
the cutoff insensitivity is preserved at higher-orders in the $1/c$
expansion, as to our knowledge no rigorously-derived relativistic equations
of motion exist in the literature. For a single particle, we see that only
at $\mathcal{O}\!\left( 1/c^{4}\right) $ does the (dipole approximation to
the) $\mathbf{A}\!\left( \mathbf{x}\right) ^{2}$ interaction play a role,
but at the same order multipole terms in the $\mathbf{p\cdot A}\!\left( 
\mathbf{x}\right) $ interaction must also be included for consistency. The
standard results show that inclusion of the $\mathbf{A}\!\left( \mathbf{x}%
\right) ^{2}$ interaction plus dipole approximation results in pathological
equations of motion. It will be very interesting to see in detail how
including the $\mathcal{O}\!\left( 1/c^{4}\right) $ multipole terms in $%
\mathbf{p\cdot A}\!\left( \mathbf{x}\right) $ might resolve these
pathologies and, in particular, whether there will be runaway-free behavior
for any cutoff (i.e., full particle-structure insensitivity) or if a finite
(or bounded) cutoff will also be required.

In conclusion, our results show that a $1/c$ effective-theory expansion
provides useful new insights into charged-particle backreaction, and
provides a systematic and consistent framework for extending to
higher-order. This is a first step towards the goal of a consistent
perturbative approach to nonequilibrium relativistic electrodynamics for
charged particle motion, considered within a nonequilibrium yet Hamiltonian
framework that incorporates a well-defined description of stochastic noise.
For instance, these results can be applied to analysis using master-equation \cite{Dipole} and influence-functional \cite{Anastopoulos98} formalisms.
Future work within this framework should also incorporate spin degrees of freedom.
Relativistic effects like particle creation, however, will probably be more
naturally described by describing the charged particles with the Dirac
field. The advantage of this formalism, in contrast, is that it more
naturally describes particle trajectories.

\section*{Acknowledgments}

\noindent We would like to thank Albert Roura for discussions of his related
work on the radiation-reaction problem. C.H.F. and B.L.H. are supported
partially by the National Science Foundation under grant PHY-0801368 to the
University of Maryland, and P.R.J. from the Research Corporation for Science
Advancement.

\appendix


\section{Quantum Brownian Motion}

\label{sec:QBM} We begin our discussion with the Quantum Brownian Motion
(QBM) Lagrangian which we have adapted in form and notation to better mirror
the problem of backreaction in the electromagnetic field. This Lagrangian
describes a quantum system bilinearly coupled to a bosonic bath of harmonic
oscillators and is traditionally used to model ordinary motional damping in
quantum mechanics. 
\begin{align}
\mathcal{L}_\mathrm{QBM} &= \mathcal{L}_\mathrm{sys} + \mathcal{L}_\mathrm{%
int} + \mathcal{L}_\mathrm{env} \, , \\
\mathcal{L}_\mathrm{sys} &\equiv \frac{1}{2} m \dot{\mathbf{x}}^2 - U(%
\mathbf{x}) \, , \\
\mathcal{L}_\mathrm{int} &= e \, \mathbf{x} \cdot \dot{\mathbf{Q}} \, , \\
\mathcal{L}_\mathrm{env} &= \int_{0}^{\infty} \!\!\! dk \, \frac{1}{2}
\left\{ \dot{\mathbf{q}}_k^2 - \omega_k^2 \, \mathbf{q}_k^2 \right\} \, ,
\end{align}
where $\mathbf{x}$ denotes the system position, $\mathbf{q}_k$ denote the
field-mode ``positions'', and $\mathbf{Q}$ is the collective field operator 
\begin{align}
\mathbf{Q} &\equiv \int_{0}^{\infty} \!\!\! dk \, g_k \, \mathbf{q}_k \, .
\end{align}
The system + environment Hamiltonian is then given by 
\begin{align}
\mathbf{H}_\mathrm{QBM} &= \mathbf{H}_\mathrm{sys} + \int_{0}^{\infty}
\!\!\! dk \, \frac{1}{2} \left\{ \left(\boldsymbol{\pi}_k - e g_k \mathbf{x}
\right)^2 + \omega_k^2 \, \mathbf{q}_k^2 \right\} ,  \label{eq:HQBM} \\
\mathbf{H}_\mathrm{sys} &\equiv \frac{\mathbf{p}^2}{2m} + U(\mathbf{x}) \, ,
\end{align}
where, as determined by the gauge of our Lagrangian, $\mathbf{p}$ is the
system momentum conjugate to $\mathbf{x}$ and $\boldsymbol{\pi}$ is the
field ``momentum'' conjugate to $\mathbf{q}$.

Note that for $m\geq 0$ and $U(\mathbf{x})$ sufficiently well behaved,
Hamiltonian \eqref{eq:HQBM} is bounded from below in its energy spectrum.
Therefore, under these conditions runaway solutions will not occur when the
environment is initially described by a thermal state. 

Additionally note that the \textquotedblleft bare\textquotedblright\ system
potential in Eq.~\eqref{eq:HQBM} is given by 
\begin{equation}
U_{\mathrm{bare}}(\mathbf{x})=U(\mathbf{x})+\left( e^{2}\!\int_{0}^{\infty
}\!\!\!dk\,\frac{g_{k}^{2}}{2}\right) \mathbf{x}^{2}\,,  \label{eq:HQBM2}
\end{equation}%
and that the system + environment Hamiltonian can also be expressed as 
\begin{equation*}
\mathbf{H}_{\mathrm{QBM}}=\mathbf{H}_{\mathrm{sys}}^{\mathrm{bare}}-e\,%
\mathbf{x}\cdot \boldsymbol{\mbox{\Large$\pi$}}+\int_{0}^{\infty }\!\!\!dk\,%
\frac{1}{2}\left\{ \boldsymbol{\pi }_{k}^{2}+\omega _{k}^{2}\,\mathbf{q}%
_{k}^{2}\right\} \,,
\end{equation*}%
in terms of the collective field operator 
\begin{equation*}
\boldsymbol{\mbox{\Large$\pi$}}\equiv \int_{0}^{\infty }\!\!\!dk\,g_{k}\,%
\boldsymbol{\pi }_{k}\,.
\end{equation*}

The resulting Heisenberg equations of motion then dictate that the system is
driven by the field 
\begin{align}
\dot{\mathbf{x}} &= \frac{\mathbf{p}}{m} \, , \\
\dot{\mathbf{p}} &= -\boldsymbol{\nabla}U_\mathrm{bare}(\mathbf{x}) + e \, 
\boldsymbol{\mbox{\Large$\pi$}} \, ,  \label{eq:dp(t)}
\end{align}
whereas the field modes are driven by the system 
\begin{align}
\dot{\mathbf{q}}_k &= \boldsymbol{\pi}_k - e g_k \mathbf{x} \, , \\
\dot{\boldsymbol{\pi}}_k &= -\omega_k^2 \, \mathbf{q}_k \, .
\end{align}

Solving for the field-mode evolution as driven by the system, we obtain the
homogeneous + driven solution 
\begin{align}
\boldsymbol{\pi}_k(t) &= \boldsymbol{\xi}_k(t) + e \, g_k \omega_k^2 \, (G_k
* \mathbf{x})(t) \, , \\
\boldsymbol{\xi}_k(t) &\equiv \dot{G}_k(t) \, \boldsymbol{\pi}_k(0) + \ddot{G%
}_k(t) \, \mathbf{q}_k(0) \, , \\
G_k(t) &\equiv \frac{\sin(\omega_k t)}{\omega_k} \, ,
\end{align}
where the $*$ product denotes the Laplace convolution 
\begin{align}
(A*B)(t) &\equiv \int_0^t \!\! dt^{\prime }\, A(t \!-\! t^{\prime }) \,
B(t^{\prime }) \, .
\end{align}
The time-evolving field operator is then given by 
\begin{align}
\boldsymbol{\mbox{\Large$\pi$}}(t) &= \underbrace{\boldsymbol{\xi}(t)}_%
\mathrm{noise} - \underbrace{2 \, e \, (\mu * \dot{\mathbf{x}})(t)}_\mathrm{%
dissipation} \, , \\
\boldsymbol{\xi}(t) &\equiv \int_{0}^{\infty} \!\!\! dk \, g_k \, 
\boldsymbol{\xi}_k(t) \, , \\
\mu(t) &\equiv - \int_{0}^{\infty} \!\!\! dk \, \frac{g_k^2 \omega_k^2}{2}
G_k(t) \, ,
\end{align}
where $\mu(t,t^{\prime }) = \mu(t - t^{\prime })$ is the stationary
dissipation kernel and $\boldsymbol{\xi}(t)$ is a Gaussian stochastic
process for the initial conditions we assume: a factorized state of the
system and environment, with the environment in a thermal state.

Next we introduce the related damping kernel 
\begin{align}
\mu(t,t^{\prime }) &= -\frac{\partial}{\partial t^{\prime }}
\gamma(t,t^{\prime }) \, ,  \label{eq:gammamu} \\
\gamma(t,t^{\prime }) & \equiv \int_{0}^{\infty} \!\!\! dk \, \frac{g_{k}^{2}%
}{2} \cos[\omega_{k}(t\!-\!t^{\prime })] \, ,
\end{align}
which is necessarily positive definite and independent of the (factorized)
initial state of the environment. The backreaction can then be expressed 
\begin{equation}
\underbrace{(\mu * \mathbf{x})(t) }_\mathrm{dissipation} = \underbrace{%
(\gamma * \dot{\mathbf{x}})(t) }_\mathrm{damping} + \underbrace{\gamma(t,0)\,%
\mathbf{x}(0)}_{\mathrm{slip}} \, - \!\!\!\! \underbrace{\gamma(t,t)\,%
\mathbf{x}(t)}_{\mathrm{renormalization}} ,
\end{equation}
in terms of the positive-definite damping and where we have labeled the
terms corresponding to the renormalization and initial short-time slip
dynamics associated with the factorization of the initial state (see Sec.~%
\ref{sec:factor}).

Substituting our field solutions into the system equations of motion, we
obtain the quantum Langevin equation 
\begin{equation*}
m\ddot{\mathbf{x}}(t)+2\,e^{2}\,(\gamma \ast \dot{\mathbf{x}})(t)+%
\boldsymbol{\nabla }U(\mathbf{x})=e\,\boldsymbol{\xi }(t)-e^{2}\,\gamma (t)\,%
\mathbf{x}(0)\,,
\end{equation*}%
which reduces to 
\begin{equation}
m\ddot{\mathbf{x}}(t)+2\,e^{2}\,(\gamma \ast \dot{\mathbf{x}})(t)+%
\boldsymbol{\nabla }U(\mathbf{x})=e\,\boldsymbol{\xi }(t)\,,
\label{eq:QBMLangevin}
\end{equation}%
after the transient slip is taken into account.


\subsection{Ohmic Coupling and Local Damping}

Considering the damping kernel, which is given by 
\begin{align}
\gamma(t) & = \int_{0}^{\infty} \!\!\! dk \, \frac{g_{k}^{2}}{2}
\cos(\omega_{k}t) \, .
\end{align}
If assume $\omega_k = c \, k$ and $g_k \approx g$ up to some high-frequency
cutoff $\Lambda$, then we may evaluate the integral as 
\begin{align}
\gamma(t) & = \frac{g^{2}}{2c} \int_{0}^{\Lambda} \!\!\! d\omega \,
\cos(\omega t) = \frac{g^{2}}{2c} \frac{\sin(\Lambda t)}{t} \, .
\end{align}
The damping kernel may then be expressed 
\begin{align}
\gamma(t) &= 2 \, \gamma_0 \, \delta_\Lambda(t) \, , \\
\gamma_0 & \equiv \frac{\pi g^{2}}{4c} \, , \\
\delta_\Lambda(t) & \equiv \frac{\sin(\Lambda t)}{\pi t} \, ,
\end{align}
in terms of the Dirac delta $\delta_\Lambda(t)$. In the high-frequency
limit, the damping contribution to the Langevin equation becomes 
\begin{align}
\lim_{\Lambda \to \infty} (\gamma * \dot{\mathbf{x}})(t) &= \gamma_0 \, \dot{%
\mathbf{x}}(t) \, ,
\end{align}
or local damping.


\subsection{Renormalization}

For Ohmic coupling or local damping the quantum Langevin equation described
by Eq.~\eqref{eq:QBMLangevin} is phenomenological, in the sense that its
various parameters correspond to the physically measurable parameters at low
energy. Assuming the Langevin equation to be phenomenological, note the bare
system potential in Hamiltonian perspective \eqref{eq:HQBM2} \& %
\eqref{eq:dp(t)} as compared to the phenomenological system potential $U(%
\mathbf{x})$ is 
\begin{equation*}
U_{\mathrm{bare}}(\mathbf{x})=U(\mathbf{x})+2\,e^{2}\,\gamma (0)\,\mathbf{x}%
^{2}\,,
\end{equation*}%
where $\gamma (0)=\frac{g^{2}}{2c}\Lambda $ for local damping with a hard
cutoff regulator. The renormalization is a quadratic term, regardless of
whether or not the original model contained such a term. 
The QBM model typically proceeds from an $\mathbf{x}\cdot \mathbf{Q}$
interaction, where this renormalization does not naturally result from the
Lagrangian theory.


\subsection{Factorized Initial Conditions}

\label{sec:factor}

If the operator noise $\boldsymbol{\xi}(t)$ in our Langevin equation is to
be sampled from a thermal distribution which is (initially) statistically
independent from the system, then the initial state of the system and
environment must be a product state of the form $\boldsymbol{\rho}_{\mathrm{%
sys+env}}=\boldsymbol{\rho}_{\mathrm{sys}} \otimes \boldsymbol{\rho}_{%
\mathrm{env}}$ or a product of marginal phase-space distributions in the
classical regime, and with the environment initially in equilibrium. This is
an important simplification in our (and most other, e.g., \cite%
{Feynman63,CaldeiraLeggett83,HPZ92}) analyses of the nonequilibrium dynamics
of open systems.

The consequence of assuming an initially uncorrelated system and environment
must be carefully examined when studying radiation reaction, however, since
acausal behaviors arise during the same very short time scale where the
unphysical nature of a factorized state is relevant. It is therefore an
important aspect of our analysis that we are also able to apply recent
results \cite{QBM} showing that for classical or high-temperature
electromagnetic noise ($\hbar \, \omega_\mathrm{sys} \ll k_\mathrm{B} T$ in
Eq.~\eqref{eq4:FDR}) the initial evolution of factorized states (or
distributions) quickly leads to physical, dressed particle states without
reintroducing the pathologies or instabilities in the dynamics that our
analysis is intended to avoid. In the semiclassical or quantum regime, use
of properly-correlated initial states can mitigate the unphysical aspects of
assuming initially factorized states entirely, without otherwise spoiling
the results in this paper \cite{Correlations}.


\subsubsection{The Slip}

The transient \emph{slip} in our Langevin equation is an initial-time
pathology associated with vanishing correlations in the factorized initial
conditions despite non-vanishing interaction strength between the system and
field. In addition to the slip, there is a diffusive initial-time pathology,
called \emph{jolts}, which arise from correlation with the (quantum)
zero-point fluctuations of the environment. 
The slip in particular was thoroughly analyzed in \cite{QBM}, where it was
pointed out to generate the linear dynamical map 
\begin{align}
\boldsymbol{\rho} & \to e^{+\imath 2 e^2 \gamma_0 \mathbf{x}^2} \, 
\boldsymbol{\rho} \, e^{-\imath 2 e^2 \gamma_0 \mathbf{x}^2} \, ,
\end{align}
which maps states in a unitary fashion and preserves all kinematic moment
invariants \cite{Dragt92}, including the uncertainty function. Therefore one
can identify the post-slip state as a ``renormalized'' initial state which
is more properly correlated with the environment and the pre-slip state as a
``bare'' initial state. If one only considers the classical regime, then
jolting is not severe due to the lack of zero-point fluctuations in the
environment. Moreover, for a classical zero-temperature environment there is
no noise causing any diffusion. Thus for this case one only needs to
consider the renormalized initial states, effectively discarding the slip
term entirely.


\subsection{The Fluctuation-Dissipation Relation}

\label{sec:FDR}

The Gaussian process $\boldsymbol{\xi}(t)$ has the noise kernel 
\begin{equation}
\nu(t,t^{\prime })=\frac{1}{2}\left\langle \{\boldsymbol{\xi}(t),\boldsymbol{%
\xi}(t^{\prime })\}\right\rangle _{\boldsymbol{\xi}}=\nu(t\!-\!t^{\prime
})\,,  \label{eq:anti-commutator}
\end{equation}
which is stationary and positive definite for any stationary initial state
of the environment. For an equilibrium initial state of the environment the
noise kernel is related to damping kernel by the (quantum)
fluctuation-dissipation relation (FDR) 
\begin{equation}
\tilde{\nu}(\omega)=\tilde{\gamma}(\omega) \, \hbar \omega \coth\!\left( 
\frac{\hbar \omega}{2 k_\mathrm{B} T}\right) \,,  \label{eq4:FDR}
\end{equation}
where $\tilde{\gamma}(\omega)=\int_{-\infty}^{+\infty}\!dt\,e^{-\imath
\omega t}\,\gamma(t)$. Essentially, the damping kernel and temperature
completely characterize Gaussian, thermal noise.

Also note that as the coupling and environment are dynamically linear, the
damping kernel, being determined by the commutator, is independent of the
state of the environment and it is the same whether in the classical or
quantum regimes. In the classical regime we have the limit 
\begin{equation}
\lim_{\hbar \to 0} \tilde{\nu}(\omega)=\tilde{\gamma}(\omega) \, 2 k_\mathrm{%
B} T \,,
\end{equation}
and the classical fluctuations vanish in the zero temperature limit. In this
limit (the classical vacuum) we can neglect the stochastic process $%
\boldsymbol{\xi}(t)$.

In the quantum regime, the anti-commutator expectation value %
\eqref{eq:anti-commutator} is not sufficient to describe the statistics of
the operator-valued stochastic process $\boldsymbol{\xi}(t)$. One
additionally requires the commutator expectation value 
\begin{equation}
\mu(t,t^{\prime })=\frac{1}{2\imath \hbar}\left\langle [\boldsymbol{\xi}(t),%
\boldsymbol{\xi}(t^{\prime })]\right\rangle_{\boldsymbol{\xi}}= \mu(t \!-\!
t^{\prime }) \, ,
\end{equation}
which is given by the dissipation kernel, a state-independent quantity. Here
the dissipation kernel is not generating backreaction, but consistent time
evolution for the non-commuting stochastic process. The full quantum
correlation is therefore given by 
\begin{align}
\alpha(t,t^{\prime }) &\equiv \left\langle \underline{\boldsymbol{%
\mbox{\Large$\pi$}}}(t) \, \underline{\boldsymbol{\mbox{\Large$\pi$}}}%
(t^{\prime }) \right\rangle_\mathrm{env} = \left\langle \boldsymbol{\xi}(t)
\, \boldsymbol{\xi}(t^{\prime }) \right\rangle_{\boldsymbol{\xi}} \, , \\
& = \nu(t \!-\! t^{\prime }) + \imath \, \hbar \, \mu( t \!-\! t^{\prime })
\, ,
\end{align}
where $\underline{\boldsymbol{\mbox{\Large$\pi$}}}(t)$ denotes the
interaction-picture or Dirac-picture field operator and not the
Heisenberg-picture field operator which we have already denoted $\boldsymbol{%
\mbox{\Large$\pi$}}(t)$.


\subsection{Stability Analysis}

\label{sec:QBMstable} We will now show that the dynamics of the system are
dissipative and stable under the very same conditions for which the system +
environment Hamiltonian \eqref{eq:HQBM} has a lower bound in its energy
spectrum. Let us denote the canonical system Hamiltonian 
\begin{align}
\mathbf{H}_\mathrm{sys} &\equiv \frac{\mathbf{p}^2}{2m} + U(\mathbf{x}) \, .
\end{align}
One may then calculate an energy constraint from either the Heisenberg
equations of motion for $\mathbf{H}_\mathrm{sys}(t)$ or by integrating the
(classical) Langevin equation \eqref{eq:QBMLangevin} along with velocity.
Accounting for the slip in our initial state, which only produces a finite
change in energy, we obtain the relation 
\begin{align}
\mathbf{H}_\mathrm{sys}(t) &= \mathbf{H}_\mathrm{sys}(0) - \mathbf{H}%
_\gamma(t) + \mathbf{H}_\xi(t) \, ,  \label{eq:stability}
\end{align}
in terms of the energy lost to damping $\mathbf{H}_\gamma(t)$ and energy
generated by noise $\mathbf{H}_\xi(t)$ 
\begin{align}
\mathbf{H}_\gamma(t) & = e^2 \!\! \int_0^t \!\!\! dt^{\prime }\!\! \int_0^t
\!\!\! dt^{\prime \prime }\, \gamma(t^{\prime },t^{\prime \prime }) \, \frac{%
1}{2} \left\{ \dot{\mathbf{x}}(t^{\prime }) \cdot \dot{\mathbf{x}}(t^{\prime
\prime }) + \dot{\mathbf{x}}(t^{\prime \prime }) \cdot \dot{\mathbf{x}}%
(t^{\prime }) \right\} , \\
\mathbf{H}_\xi(t) & = e \! \int_0^t \!\!\! dt^{\prime }\, \frac{1}{2}
\left\{ \boldsymbol{\xi}(t^{\prime }) \cdot \dot{\mathbf{x}}(t^{\prime }) + 
\dot{\mathbf{x}}(t^{\prime }) \cdot \boldsymbol{\xi}(t^{\prime }) \right\} .
\end{align}
The contribution from damping is a negative quantity as the damping kernel
is a positive-definite kernel in a quadratic form. The noise is random and
may drive the system erratically, but the damping may only remove energy
from the system (and deliver it to the environment and interaction).
Therefore it is imperative that $\mathbf{H}_\mathrm{sys}$ have a lower bound
in its energy spectrum. For our model, this necessarily implies that the
system + environment Hamiltonian \eqref{eq:HQBM} also has a lower bound in
its energy spectrum. If this is the case then true runaway motion cannot
occur. In the classical-vacuum limit, energy is continually siphoned from $%
\mathbf{H}_\mathrm{sys}(t)$ until all motion ceases.

Locally-damped energy is additionally simplified to 
\begin{align}
\dot{\mathbf{H}}_\gamma(t) &= 2 \, e^2 \, \gamma_0 \, \dot{\mathbf{x}}(t)^2
\, ,
\end{align}
which monotonically dissipates energy in time. Nonlocal damping can produce
an instantaneous increase in system energy, though the cumulative effect is
always dissipative.


\section{Explicit Calculation of Driven Quantum Field}

\label{sec:Explicit} Most simply let us consider the field degrees of
freedom as driven by one particle, using Hamiltonian \eqref{eq:HEQED}. 
\begin{align}
\dot{\mathbf{a}}_{\mathbf{k},\boldsymbol{\epsilon}_k} &= - \imath \,
\omega_k \, {\mathbf{a}}_{\mathbf{k},\boldsymbol{\epsilon}_k} - \imath
\sum_j \frac{\frac{e_j}{2m_j}}{\sqrt{2 \varepsilon_0 \omega_k}} \left\{
e^{-\imath \, \mathbf{k} \cdot \mathbf{x}_j} , (\boldsymbol{\epsilon}_k
\cdot \mathbf{p}_j) \right\} \, .
\end{align}
The driven solutions of each field mode are therefore given by 
\begin{align}
& \mathbf{a}_{\mathbf{k},\boldsymbol{\epsilon}_k}\!(t) = e^{- \imath
\omega_k t} \, {\mathbf{a}}_{\mathbf{k},\boldsymbol{\epsilon}_k}\!(0) \\
& - \imath \sum_j \frac{\frac{e_j}{2m_j}}{\sqrt{2 \varepsilon_0 \omega_k}}
\int_0^t \!\! dt^{\prime - \imath \omega_k (t-t^{\prime })} \left\{
e^{-\imath \, \mathbf{k} \cdot \mathbf{x}_j(t^{\prime })} , \boldsymbol{%
\epsilon}_k \cdot \mathbf{p}_j(t^{\prime }) \right\} \, .  \notag
\end{align}
In calculating $\mathbf{A}(\mathbf{x}_i)$ \eqref{eq:Ak2} we require
evaluation of the field modes at the system location 
\begin{align}
e^{+\imath \, \mathbf{k} \cdot \mathbf{x}_i(t)} \, \mathbf{a}_{\mathbf{k}, 
\boldsymbol{\epsilon}_k}(t) &= \frac{1}{2} \left\{ e^{+\imath \, \mathbf{k}
\cdot \mathbf{x}_i(t)} , \mathbf{a}_{\mathbf{k},\boldsymbol{\epsilon}%
_k}\!(t) \right\} \, ,
\end{align}
which we have placed into symmetric form by commutativity of the system and
field operators. We now must consider the driven mode 
\begin{widetext}
\begin{align}
& e^{+\imath \, \mathbf{k} \cdot \mathbf{x}_i(t)} \, \mathbf{a}_{\mathbf{k},\boldsymbol{\epsilon}_k}(t)
= \frac{1}{2} \left\{ e^{- \imath (\omega_k t - \mathbf{k} \cdot \mathbf{x}_i(t))} , {\mathbf{a}}_{\mathbf{k},\boldsymbol{\epsilon}_k}\!(0) \right\}
- \imath \sum_j \frac{\frac{e_j}{4m_j}}{\sqrt{2 \varepsilon_0 \omega_k}}  \int_0^t \!\! dt' \left\{ e^{- \imath (\omega_k t - \mathbf{k} \cdot \mathbf{x}_i(t))} , \left\{ e^{+ \imath (\omega_k t' - \mathbf{k} \cdot \mathbf{x}_j(t'))} , \boldsymbol{\epsilon}_k \cdot \mathbf{p}_j(t') \right\} \right\} ,
\end{align}
\end{widetext}
which we have also placed into a manifestly Hermitian form. Note that any
function $f[\mathbf{k} \cdot \mathbf{x}(t^{\prime })]$ commutes with $%
\boldsymbol{\epsilon}_k \cdot \mathbf{p}(t^{\prime })$ given the Coulomb
gauge constraint $\mathbf{k} \cdot \boldsymbol{\epsilon}_k$. Therefore we
may move $\boldsymbol{\epsilon}_k \cdot \mathbf{p}(t^{\prime })$ to the most
suitable side of $e^{+ \imath \mathbf{k} \cdot \mathbf{x}(t^{\prime })}$.
For the velocity source, one can first decompose the velocity into momentum $%
\boldsymbol{\epsilon}_k \cdot \mathbf{p}(t^{\prime })$ and field $%
\boldsymbol{\epsilon}_k \cdot \mathbf{A}[\mathbf{x}(t^{\prime })]$ and then
note that by the previous argument both terms commute with any function $f[%
\mathbf{k} \cdot \mathbf{x}(t^{\prime })]$. For the position source,
commutativity follows trivially.

In any case, one can produce relations such as \eqref{eq:A(p)} by
substitution of the above expression into Eq.~$\eqref{eq:Ak2}$: 
\begin{equation}
\mathbf{A}(\mathbf{x}_{i},t)=\boldsymbol{\xi }_{i}^{A}(t)-\sum_{j}\frac{e_{j}%
}{m_j}\left\{ (\boldsymbol{\mu }_{ij}^{A}\ast {\mathbf{p}}_{j})(t)+(%
\boldsymbol{\mu}_{ij}^{A}\ast \dot{\mathbf{p}}_{j})^{\dagger }(t)\right\} \,,
\end{equation}
where the dissipation convolution is given by 
\begin{align}
(\boldsymbol{\mu }_{ij}^{A}\ast {\mathbf{p}}_{j})(t) & \equiv \int_0^t \!\!
dt^{\prime }\, \boldsymbol{\mu }_{ij}^{A}(t,t^{\prime }) \, {\mathbf{p}}%
_{j}(t^{\prime }) \, ,
\end{align}
and with the noise and dissipation kernel microscopically determined to be 
\begin{widetext}
\begin{align}
\boldsymbol{\xi }_{i}^{A}(t) &\equiv \frac{1}{(2\pi )^{3/2}}\int \!d^{3}k \sum_{\boldsymbol{\epsilon }_{k}}
\frac{\boldsymbol{\epsilon }_{k}}{\sqrt{2\varepsilon _{0}\omega _{k}}} \frac{1}{2}
\left( \left\{ e^{- \imath (\omega_k t - \mathbf{k} \cdot \mathbf{x}_i(t))} , {\mathbf{a}}_{\mathbf{k},\boldsymbol{\epsilon}_k}\!(0) \right\}
+ \left\{ e^{+ \imath (\omega_k t - \mathbf{k} \cdot \mathbf{x}_i(t))} , {\mathbf{a}}_{\mathbf{k},\boldsymbol{\epsilon}_k}^{\dagger}\!(0) \right\} \right) \, , \\
\boldsymbol{\mu }_{ij}^{A}(t,t') & \equiv
\frac{1}{(2\pi )^{3/2}}\int \!d^{3}k \sum_{\boldsymbol{\epsilon }_{k}}
\frac{\boldsymbol{\epsilon }_{k} \, \boldsymbol{\epsilon}_k^\mathrm{T}}{2\varepsilon _{0}\omega _{k}}
\frac{\imath}{2} \left\{ e^{- \imath (\omega_k t - \mathbf{k} \cdot \mathbf{x}_i(t))} \, e^{+ \imath (\omega_k t' - \mathbf{k} \cdot \mathbf{x}_j(t'))}
- e^{+ \imath (\omega_k t - \mathbf{k} \cdot \mathbf{x}_i(t))} \, e^{- \imath (\omega_k t' - \mathbf{k} \cdot \mathbf{x}_j(t'))} \right\} \, , \label{eq:muDEF} \\
\boldsymbol{\mu }_{ij}^{A}(t,t') &=
\frac{1}{(2\pi )^{3/2}}\int \!d^{3}k \sum_{\boldsymbol{\epsilon }_{k}}
\frac{\boldsymbol{\epsilon }_{k} \, \boldsymbol{\epsilon}_k^\mathrm{T}}{2\varepsilon _{0}\omega _{k}}
\sin\!\left[ \omega_k (t\!-\!t') - \mathbf{k} \cdot (\mathbf{x}_i(t)\!-\!\mathbf{x}_j(t')) + \mathcal{O}(\hbar^2 k^3) \right] \, . \label{eq:muDEF2}
\end{align}
\end{widetext}
In the classical calculation the phase factors in Eq.~\eqref{eq:muDEF}
commute, however in the quantum-relativistic regime they give rise
nontrivial interference terms as compared to the approximate %
\eqref{eq:muDEF2}. The $\mathcal{O}(k)$ phases in \eqref{eq:muDEF2}
contribute to the $1/c^3$ dissipative forces (given microscopic structure)
and $1/c^2$ nondissipative forces (given multiple particles). These phases
are discarded in the dipole approximation. The next order $\mathcal{O}%
(\hbar^2 k^3)$ quantum phase corrections are determined by two-time
commutators of the system trajectories and are negligible to the order we
work at.

By $\mathcal{O}(\hbar)$ we only mean to keep track of the fact that these
corrections contain commutator dependence, and vanish for the classical
equations of motion. Planck's constant is not an expansion parameter which
we consider in this work. The effect of the $\mathcal{O}(k^n)$ or $\mathcal{O%
}(1/c^n)$ phase corrections admit a simple dimensional analysis, as only the
full-time trajectories $\mathbf{x}(t^{\prime })$ are input into these
highly-oscillatory integrals. Dimensionless corrections can therefore only
be formed by derivatives of $\mathbf{x}(t)$ and factors of $1/c$, though not
necessarily $v/c$. Essentially the $1/c$ expansion appears to be a kind of
multipole expansion here, which generalizes the usual dipole limit.

From Eq.~\eqref{eq:muDEF} the positive-definite and Hermitian damping kernel
is then given by 
\begin{widetext}
\begin{align}
\boldsymbol{\gamma }_{ij}^{A}(t,t') & \equiv
\frac{1}{(2\pi )^{3/2}}\int \!d^{3}k \sum_{\boldsymbol{\epsilon }_{k}}
\frac{\boldsymbol{\epsilon }_{k} \, \boldsymbol{\epsilon}_k^\mathrm{T}}{2\varepsilon _{0}\omega _{k}^2}
\frac{1}{2} \left\{ e^{- \imath (\omega_k t - \mathbf{k} \cdot \mathbf{x}_i(t))} \, e^{+ \imath (\omega_k t' - \mathbf{k} \cdot \mathbf{x}_j(t'))}
+ e^{+ \imath (\omega_k t - \mathbf{k} \cdot \mathbf{x}_i(t))} \, e^{- \imath (\omega_k t' - \mathbf{k} \cdot \mathbf{x}_j(t'))} \right\} \, , \label{eq:gammaDEF} \\
\boldsymbol{\gamma }_{ij}^{A}(t,t') &=
\frac{1}{(2\pi )^{3/2}}\int \!d^{3}k \sum_{\boldsymbol{\epsilon }_{k}}
\frac{\boldsymbol{\epsilon }_{k} \, \boldsymbol{\epsilon}_k^\mathrm{T}}{2\varepsilon _{0}\omega _{k}^2}
\cos\!\left[ \omega_k (t\!-\!t') - \mathbf{k} \cdot (\mathbf{x}_i(t)\!-\!\mathbf{x}_j(t')) + \mathcal{O}(\hbar k^2) \right] \, . \label{eq:gammaDEF2}
\end{align}
\end{widetext}
We finally note the relation 
\begin{align}
\boldsymbol{\mu }_{ij}^{A}(t,t^{\prime }) &= -\frac{d}{dt^{\prime }} 
\boldsymbol{\gamma }_{ij}^{A}(t,t^{\prime }) + \mathcal{O}\left( \frac{v}{c}
\mu^A \right) \, ,
\end{align}
and so, to the order we consider, we may integrate by parts the dissipation
kernel into the damping kernel. The relativistic correction here does not
precisely involve the dissipation kernel, but a kernel of with the same
parameter scaling and frequency sensitivity.

The operator noise processes are also more complicated quantum mechanically,
as even for a Gaussian state of the environment the noise cumulants cannot
be evaluated exactly. However, to lowest order in $1/c$, with the $\mathcal{O%
}(k)$ phases treated as quasistationary, the noise processes are dual to the 
$1/c^3$ damping, with which they satisfy a fluctuation-dissipation relation.

\bibliographystyle{apsrev4-1}
\bibliography{bib}

\end{document}